\documentclass[vecphys]{svmult}

\usepackage{makeidx}         % allows index generation
\usepackage{graphicx}        % standard LaTeX graphics tool
\usepackage{multicol}        % used for the two-column index
\usepackage[bottom]{footmisc}% places footnotes at page bottom
\makeindex             % used for the subject index
\newcommand{\be}{\begin{equation}}
\newcommand{\ee}{\end{equation}}
\newcommand{\bea}{\begin{eqnarray}}
\newcommand{\eea}{\end{eqnarray}}
\newcommand{\ba}{\begin{array}}
\newcommand{\ea}{\end{array}}
\newcommand{\bc}{\begin{center}}
\newcommand{\ec}{\end{center}}
\newcommand{\Dslash}{D\hspace*{-0.23cm}/\,}

\begin{document}

\title*{Effective Theories of Dense and Very Dense Matter}
\author{Thomas Sch\"afer\inst{1}}
\institute{Department of Physics,
     North Carolina State University,
     Raleigh, NC 27695
\texttt{thomas\_schaefer@ncsu.edu}}
\maketitle

%%%%%%%%%%%%%%%%%%%%%%%%%%%%%%%%%%%%%%%%%%%%%%%%%%%%%%%%%%%%%%%%%%%%
\section{Introduction}
\label{sec_intro}
%%%%%%%%%%%%%%%%%%%%%%%%%%%%%%%%%%%%%%%%%%%%%%%%%%%%%%%%%%%%%%%%%%%%

 The exploration of the phase diagram of dense baryonic matter 
is an area of intense theoretical and experimental activity. 
Baryonic systems, from dilute neutron matter at low density 
to superconducting quark matter at high density, exhibit an 
enormous variety of many-body effects. Despite its simplicity 
all these phenomena are ultimately described by the lagrangian
of QCD.

 In practice it is usually very difficult to describe QCD 
many-body systems directly in terms of the QCD lagrangian, 
and even in cases where this is possible it is often not 
the most convenient and most transparent description. Instead, 
it is advantageous to employ an effective field theory (EFT)
formulated in terms of the relevant degrees of freedom. EFTs
also provide a unified description of physical systems involving
very different length scales, such as Fermi liquids in nuclear 
and atomic physics, or non-Fermi liquid gauge theories 
involving colored quarks or charged electrons. 

 In these lectures we shall discuss the many body physics 
of several effective field theories relevant to the structure 
of hadronic matter. We will concentrate on two regimes in 
the phase diagram. At low baryon density the relevant degrees
of freedom are neutrons and protons, while at very high baryon
density the degrees of freedom are quarks and gluons. These
lectures do not provide an introduction to effective field
theories (see \cite{Polchinski:1992ed,Manohar:1996cq,Kaplan:2005es}), 
nor an exhaustive treatment of many body physics (see 
\cite{Abrikosov:1963,Fetter:1971,Negele:1988vy}) or the physics
of dense quark matter (see \cite{Rajagopal:2000wf,Alford:2001dt}).

%%%%%%%%%%%%%%%%%%%%%%%%%%%%%%%%%%%%%%%%%%%%%%%%%%%%%%%%%%%%%%%%%%%%
\section{Fermi liquids}
\label{sec_fl}
\subsection{Effective field theory for non-relativistic fermions}
\label{sec_nr_eft}
%%%%%%%%%%%%%%%%%%%%%%%%%%%%%%%%%%%%%%%%%%%%%%%%%%%%%%%%%%%%%%%%%%%%

  If the relevant momenta are small neutrons and protons can 
be described as point-like non-relativistic fermions interacting 
via local forces. Effective field theories for nuclear systems
have been studied extensively over the past couple of years
\cite{Kaplan:2005es,Beane:2000fx,Bedaque:2002mn,Epelbaum:2005pn}. 
If the typical momenta are on the order of the pion mass pions 
have to be included as explicit degrees of freedoms. For simplicity 
we will consider neutrons only and focus on momenta small compared 
to $m_\pi$. The effective lagrangian is 
\be 
\label{l_4f}
{\cal L}_0 = \psi^\dagger \left( i\partial_0 +
 \frac{\nabla^2}{2m} \right) \psi 
 - \frac{C_0}{2} \left(\psi^\dagger \psi\right)^2 
 + \frac{C_2}{16} \left[ (\psi\psi)^\dagger
     (\psi \stackrel{\leftrightarrow}{\nabla}{}^{\!2}\psi) +h.c \right]
 +\ldots ,
\ee
where $m$ is the neutron mass, $C_0$ and $C_2$ are dimensionful coupling 
constants, $\stackrel{\leftrightarrow}{\nabla}=\stackrel{\rightarrow}{\nabla}
-\stackrel{\leftarrow}{\nabla}$ is a Galilei invariant derivative, and 
$\ldots$ denotes interactions with more derivatives. We have only displayed 
terms that act in the s-channel. The coupling constant are determined by 
the neutron-neutron scattering amplitude. For non-relativistic scattering 
the amplitude is related to the scattering phase shift $\delta$ by
\be
\label{delta_s}
{\cal A} = \frac{4\pi}{m} \frac{1}{p\cot\delta -ip} .
\ee
For small momenta the quantity $p\cot\delta$ can be expanded as a
Taylor series in $p$. This expansion is called the effective range 
expansion
\be
\label{ere}
 p\cot\delta = -\frac{1}{a} + \frac{1}{2}
  \sum_{n=0}^\infty {r}_n p^{2(n+1)},
\ee
where $a$ is the scattering length, and $r_0$ is the effective range. 
The situation is simplest if the scattering length is small. In this 
case the scattering amplitude has a perturbative expansion in $C_i$. 
At tree level 
\be
\label{C_i_ere}
C_0 = \frac{4\pi a}{m}, \hspace{0.5cm} 
C_2 = C_0 \frac{a r_0}{2}  .
\ee
However, there are many systems of physical interest in which the 
scattering length is not small. This happens whenever there is 
a two-body bound state with a very small binding energy, or if 
the two-body system is very close to forming a bound state. For 
neutrons $a_{nn}=-17$ fm, much larger than typical strong interaction
length scales. 

 If the scattering length is large then loop diagrams with the 
leading order interaction $C_0(\psi^\dagger\psi)^2$ have to 
be resummed. The one-loop correction involves the loop integral
\bea
L(E) &=&  i\int \frac{d^{d+1} q}{ (2\pi)^{d+1}}\, 
   \frac{1}
        {(E/2 + q_0 - \vec{q}^2/(2m)+ i\epsilon)
         (E/2 - q_0 - \vec{q}^2/(2m)+ i\epsilon)}
   \nonumber\\
 &=& \int \frac{d^dq}{(2\pi)^{d}}\,
    \frac{1}{E -\vec{q}^2/m + i\epsilon}
\nonumber\\
&=& -  \frac{m}{(4\pi)^{d/2}} \,
    \Gamma\left(\frac{2-d}{2}\right)
    (-mE-i\epsilon)^{\frac{d-2}{2}} ,
\label{nn_loop}
\eea
where $E$ is the center-of-mass energy.
We have regularized the integral by analytic continuation to $d+1$ 
dimensions. In order to define the theory we have to specify a 
subtraction scheme. Here, we will employ the modified minimal 
subtraction $\overline{MS}$ scheme. See \cite{Kaplan:1998we} for a 
discussion of different renormalization schemes. We get
\be
\label{loop_uni}
L(E)=   \frac{m}{4\pi} \sqrt{-mE-i\epsilon}=
-\frac{m}{4\pi}\, ip  \, ,
\ee
where $p=\sqrt{mE}$ is the nucleon momentum in the center-of-momentum 
frame. It is now straightforward to sum all the bubble diagrams. The 
result is 
\be
\label{nn_sum}
{\cal A}  =  -\frac{C_{0}}{1 + iC_0(mp/4\pi)} .
\ee
Higher order corrections due to the $C_i$ terms ($i\geq 2$) can be 
treated perturbatively. The bubble sum can now be matched to the 
effective range expansion. In the $\overline{MS}$ scheme the result is 
particularly simple since equ.~(\ref{loop_uni}) only contains
the contribution from the unitarity cut. As a consequence, the 
result given in equ.~(\ref{C_i_ere}) is not modified even if 
$C_0$ is summed to all orders.

%%%%%%%%%%%%%%%%%%%%%%%%%%%%%%%%%%%%%%%%%%%%%%%%%%%%%%%%%%%%%%%%%%%%
\subsection{Dilute Fermi liquid}
\label{sec_dfl}
%%%%%%%%%%%%%%%%%%%%%%%%%%%%%%%%%%%%%%%%%%%%%%%%%%%%%%%%%%%%%%%%%%%%

The lagrangian given in equ.~(\ref{l_4f}) is invariant under the 
$U(1)$ transformation $\psi\to e^{i\phi}\psi$. The $U(1)$ symmetry 
implies that the fermion number 
\be
 N= \int d^3x\,\psi^\dagger \psi
\ee
is conserved. As a consequence, it is meaningful to study a 
system of fermions at finite density $\rho=N/V$. We will do 
this in the grand-canonical formalism. We introduce a chemical 
potential $\mu$ conjugate to the fermion number $N$ and 
study the partition function
\be 
\label{Z}
 Z(\mu,\beta) = {\rm Tr}\left[e^{-\beta(H-\mu N)}\right].
\ee
Here, $H$ is the Hamiltonian associated with ${\cal L}$ 
and $\beta=1/T$ is the inverse temperature. The trace in 
equ.~(\ref{Z}) runs over all possible states of the system. 
The average number of particles for a given chemical potential 
$\mu$ and temperature $T$ is given by $\langle N\rangle =T
(\partial \log Z)/(\partial \mu)$. At zero temperature the 
chemical potential is the energy required to add one particle 
to the system. 

%%%%%%%%%%%%%%%%%%%%%%%%%%%%%%%%%%%%%%%%%%%%%%%%%%%%%%%%%%%%%%%%%%%%
\begin{figure}[t]
\bc\includegraphics[width=10.0cm]{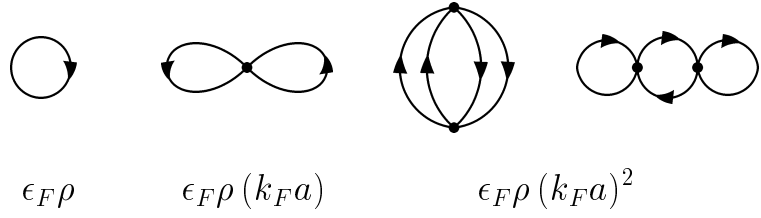}\ec
\caption{\label{fig_fl}
Leading order Feynman diagrams for the ground state 
energy of a dilute gas of fermions interacting via
a short range potential.}
\end{figure}
%%%%%%%%%%%%%%%%%%%%%%%%%%%%%%%%%%%%%%%%%%%%%%%%%%%%%%%%%%%%%%%%%%%%

 There is a formal resemblance between the partition function
equ.~(\ref{Z}) and the quantum mechanical time evolution 
operator $U=\exp(-iHt)$. In order to write the partition
function as a time evolution operator we have to identify 
$\beta\to it$ and add the term $-\mu N$ to the Hamiltonian. 
Using standard techniques we can write the time evolution
operators as a path integral \cite{Kapusta:1989,LeBellac:1996}
\be 
Z = \int D\psi D\psi^\dagger \exp\left(-\int_0^\beta d\tau 
  \int d^3x\, {\cal L}_E \right).
\ee
Here, ${\cal L}_E$ is the euclidean lagrangian
\be 
\label{l_4f_E}
{\cal L}_E = \psi^\dagger \left( \partial_\tau 
 - \mu - \frac{\nabla^2}{2m} \right) \psi 
 + \frac{C_0}{2} \left(\psi^\dagger \psi\right)^2 +\ldots \, .
\ee
The fermion fields satisfy anti-periodic boundary 
conditions $\psi(\beta)=-\psi(0)$. Equation (\ref{l_4f_E}) is 
the starting point of the imaginary time formalism in 
thermal field theory. The chemical potential simply results 
in an extra term $-\mu\psi^\dagger\psi$ in the lagrangian. 
From equ.~(\ref{l_4f_E}) we can easily read off the free 
fermion propagator 
\be
S_{\alpha\beta}^0(p) = \frac{\delta_{\alpha\beta}}
  {ip_4+\mu-\frac{\vec{p}^{\, 2}}{2m}},
\ee
where $\alpha,\beta$ are spin labels. We observe that the 
chemical potential simply shifts the four-component of the 
momentum. This implies that we have to carefully analyze the
boundary conditions in the path integral in order to fix the 
pole prescription. The correct Minkowski space propagator is
\be
\label{s_ph} 
S^0_{\alpha\beta}(p) =
 \frac{\delta_{\alpha\beta}}
 {p_0-\epsilon_p+i\delta{\rm sgn}(\epsilon_p)}
 = \delta_{\alpha\beta}\left\{
 \frac{\Theta(p-p_F)}{p_0-\epsilon_p+i\delta}+
 \frac{\Theta(p_F-p)}{p_0-\epsilon_p-i\delta}
  \right\},\nonumber
\ee
where $\epsilon_p=E_p-\mu$, $E_p=\vec{p}^{\, 2}/(2m)$ and
$\delta\to 0^+$. The quantity $p_F=\sqrt{2m\mu}$ is called 
the Fermi momentum. We will refer to the surface defined 
by the condition $|\vec{p}|=p_F$ as the Fermi 
surface. The two terms in equ.~(\ref{s_ph}) have a simple 
physical interpretation. At finite density and zero temperature
all states with momenta below the Fermi momentum are 
occupied, while all states above the Fermi momentum are
empty. The possible excitation of the system are particles
above the Fermi surface or holes below the Fermi surface,
corresponding to the first and second term in 
equ.~(\ref{s_ph}). The particle density is given by
\be
\rho = \langle\psi^\dagger\psi\rangle =
 \int \frac{d^4p}{(2\pi)^4} S^0_{\alpha\alpha}(p)
 \left. e^{ip_0\delta}\right|_{\delta\to 0^+}
 = 2\int \frac{d^3p}{(2\pi)^3}\Theta(p_F-p)
 = \frac{p_F^3}{3\pi^2}.
\ee 
Tadpole diagrams require an extra $i\delta$ prescription which 
can be derived from a careful analysis of the path integral 
representation at $\mu\neq 0$. As a first simple application we 
can compute the energy density as a function of the fermion density. 
For free fermions, we find 
\be
{\cal E} =  2\int \frac{d^3p}{(2\pi)^3}\, E_p\Theta(p_F-p)
 = \frac{3}{5}\rho\frac{p_F^2}{2m}.
\ee
We can also compute the corrections to the ground state energy due 
to the interaction $(C_0/2)(\psi^\dagger\psi)^2$. The first term is 
a two-loop diagram with one insertion of $C_0$, see Fig.~\ref{fig_fl}. 
There are two possible contractions and the spin-factor of the diagram 
is $(\delta_{\alpha\alpha}\delta_{\beta\beta}-\delta_{\alpha\beta}
\delta_{\alpha\beta})=g(g-1)$ where $g=(2s+1)$ is the degeneracy and 
$s$ is the spin of the fermions. In the following we will always
set $g=2$. The diagram is proportional to the square of the density
and we get 
\be 
\label{e1}
{\cal E}_1 = C_0\left(\frac{p_F^3}{6\pi^2}\right)^2.
\ee
We observe that the sum of the first two terms in the energy density 
can be written 
as 
\be
\label{e_pfa}
{\cal E} = \rho\, \frac{p_F^2}{2m}\left(
\frac{3}{5} + \frac{2}{3\pi}(p_Fa)+\ldots  \right),
\ee
which shows that the $C_0$ term is the first term in an expansion 
in $p_Fa$, suitable for a dilute, weakly interacting, Fermi gas. 

%%%%%%%%%%%%%%%%%%%%%%%%%%%%%%%%%%%%%%%%%%%%%%%%%%%%%%%%%%%%%%%%%%%%
\subsection{Higher order corrections}
\label{sec_kfa2}
%%%%%%%%%%%%%%%%%%%%%%%%%%%%%%%%%%%%%%%%%%%%%%%%%%%%%%%%%%%%%%%%%%%%

The expansion in $(p_Fa)$ was carried out to order $(p_Fa)^2$ by 
Huang, Lee and Yang \cite{Lee:1957,Huang:1957}. Since then, the 
accuracy was pushed to $O((p_Fa)^4\log(p_Fa))$, see \cite{Hammer:2000xg} 
for an EFT approach to this calculation. The $O((p_Fa)^2)$ calculation 
involves a few new ingredients and we shall briefly outline the main 
steps. Consider the third diagram in Fig.~\ref{fig_fl}. The contribution 
to the vacuum energy is 
\be 
{\cal E}_2 = -i\frac{C_0^2}{2}
  \int \frac{d^4q_1}{(2\pi)^4}
  \int \frac{d^4q_2}{(2\pi)^4}
  \int \frac{d^4q_3}{(2\pi)^4}
  S(q_1)S(q_2)S(q_3)S(q_1+q_2-q_3) .
\ee
We begin by performing two of the energy integrals using contour
integration. The contours can be placed in such a way that the 
two poles correspond to two particles or two holes (but not a 
particle and a hole). This allows us to write 
\be
{\cal E}_2 =  \frac{C_0^2}{2}  
    \int \frac{d^3q_1}{(2\pi)^3}
    \int \frac{d^3q_2}{(2\pi)^3} \, \theta_q^-
         \Pi_{pp} (q_1+q_2) + h.c. \,  ,
\ee
where $\theta_q^-$ is the Pauli-blocking factor corresponding 
to a pair of holes
\be
\label{th_hh}
\theta_q^- =  \theta\left(p_F-q_1\right) \,
                       \theta\left(p_F-q_2\right) \ ,
\ee
and $\Pi_{pp}$ is the one-loop particle-particle scattering 
amplitude. Since $q_{1,2}$ are on-shell we can write $\Pi_{pp}$
as a function of the center-of-mass and relative momenta $\vec{P}$ 
and $\vec{k}$ with $\vec{q}_{1,2}=\vec{P}/2\pm\vec{k}$. Note that 
because of Galilean invariance the vacuum scattering amplitude
only depends on $\vec{k}$. We find
\be
\label{fpp}
\Pi_{pp} = 
\int\!\! \frac{d^3q}{(2\pi)^3} \;
  \frac{m\theta_q^+}{\vec{k}^2-\vec{q}^2+i\epsilon} 
 = \Pi_{pp}^{vac}(k) + \frac{mp_F}{(2\pi)^2} f_{pp}(\kappa,s) \ ,
\ee
where $\theta_q^+ =\theta\left(p_F+q_1\right)\theta\left(p_F+q_2\right)$
is defined in analogy with equ.~(\ref{th_hh}). The first term on the 
RHS is the vacuum contribution and the second term is the medium 
contribution which depends on the scaled momenta $\vec{\kappa}=
\vec{k}/p_F$ and $\vec{s}=\vec{P}/(2p_F)$. The vacuum contribution 
is divergent and needs to be renormalized. In dimensional regularization
$\Pi_{pp}^{vac}$ is purely imaginary and does not contribute to the 
vacuum energy. In other regularization schemes the vacuum contributions
combines with the $O(C_0)$ graph to give the correct one-loop relation
between $C_0$ and the scattering length. 

For $s<1$ the in-medium scattering amplitude is given by
\be
\label{fpp_2}
 f_{PP}(\kappa,s) = 1+s+
   \kappa \log \left|\frac{1+s-\kappa}{1+s+\kappa}\right|
   + \frac{1-\kappa^2-s^2}{2s} \log \left|
    \frac{(1+s)^2-\kappa^2}{1-\kappa^2-s^2}\right| \ .
\ee 
The contribution to the energy density can now be determined by 
integrating equ.~(\ref{fpp_2}) over phase space. We find
\bea
\label{E_pp}
{\cal E}_2 &=& C_0^2\frac{p_F m}{4\pi^2}
  \int\! \frac{d^3P}{(2\pi)^3}\frac{d^3k}{(2\pi)^3} 
      \; \theta_k^- \; f_{PP}(\kappa,s) \nonumber \\
 &=& \rho \, \frac{p_F^2}{2m}\frac{4}{35\pi^2}
       \left(11-2\log(2)\right) \left(p_Fa\right)^2 . 
\eea
The fourth diagram in Fig.~\ref{fig_fl} involves a particle-hole
pair with zero energy and the corresponding phase space factor 
vanishes \cite{Fetter:1971}.

 The effective lagrangian can also be used to study many other 
properties of the system, such as corrections to the fermion propagator. 
Near the Fermi surface the propagator can be written as
\be
\label{s_qp}
 S_{\alpha\beta} = \frac{Z\delta_{\alpha\beta}}
    {p_0-v_F(|\vec{p}|-p_F)+i\delta{\rm sgn}(|\vec{p}|-p_F)},
\ee
where $Z$ is the wave function renormalization and $v_F=p_F/m^*$ is 
the Fermi velocity. $Z$ and $m^*$ can be worked out order by order 
in $(p_Fa)$, see \cite{Abrikosov:1963,Platter:2002yr}. The leading
order results are
\bea
\frac{m^*}{m} &=& 1-\frac{8}{15\pi^2}\left(1-7\log(2)\right)
       \left(p_Fa\right)^2 + \ldots \\
 Z^{-1} &=& 1-\frac{4}{\pi^2} \log(2)
       \left(p_Fa\right)^2 + \ldots 
\eea
The main observation is that the structure of the propagator is 
unchanged even if interactions are taken into account. The low energy 
excitations are quasi-particles and holes, and near the Fermi surface 
the lifetime of a quasi-particle is infinite. This is the basis of Fermi 
liquid theory \cite{Pines:1966,Baym:1991}.

%%%%%%%%%%%%%%%%%%%%%%%%%%%%%%%%%%%%%%%%%%%%%%%%%%%%%%%%%%%%%%%%%%%%
\begin{figure}[t]
\bc\includegraphics[width=8.0cm]{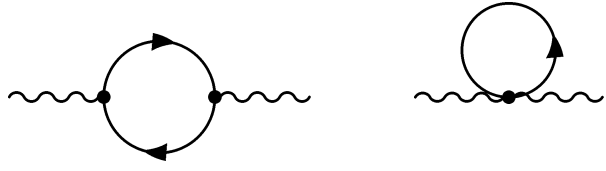}\ec
\caption{\label{fig_screen}
Leading order Feynman diagrams that contribute to the 
photon polarization function in a non-relativistic Fermi 
liquid. The tadpole diagram shown in the right panel 
only appears in the spatial part of the polarization 
tensor.}
\end{figure}
%%%%%%%%%%%%%%%%%%%%%%%%%%%%%%%%%%%%%%%%%%%%%%%%%%%%%%%%%%%%%%%%%%%%

%%%%%%%%%%%%%%%%%%%%%%%%%%%%%%%%%%%%%%%%%%%%%%%%%%%%%%%%%%%%%%%%%%%%
\subsection{Screening and damping}
\label{sec_screen}
%%%%%%%%%%%%%%%%%%%%%%%%%%%%%%%%%%%%%%%%%%%%%%%%%%%%%%%%%%%%%%%%%%%%

An important aspect of a dilute Fermi gas of charged particles is 
the response to an external electromagnetic field. We consider a 
system in which the total charge is neutralized by a homogeneous 
background (such as positive ions in a metal). The response to an 
electric field is governed by the gauge coupling $eA_0\psi^\dagger\psi$. 
The medium correction to the photon propagator is determined by the 
polarization function
\be
\Pi_{00}(q) = e^2 \int d^4x\, e^{-iqx}
 \langle \psi^\dagger \psi(0)\psi^\dagger \psi(x)
\rangle .
\ee
The one-loop contribution is given by
\be
\Pi_{00}(q) = -ie^2 \int\!\frac{d^4p}{(2\pi)^4}
 \frac{1}{q_0+p_0-\epsilon_{p+q}+
        i\delta{\rm sgn}(\epsilon_{p+q})}
 \frac{1}{p_0-\epsilon_{p}+
        i\delta{\rm sgn}(\epsilon_{p})}.
\ee
Performing the $p_0$ integral using contour integration we find
\be
 \Pi_{00}(q) = e^2\int\!\frac{d^3p}{(2\pi)^3} 
\frac{n_{p+q}-n_p}{E_{p+q}-E_p},
\ee
where we have introduced the Fermi distribution function $n_p=\Theta
(p_F-p)$. We observe that in the limit $\vec{q}\to 0$ the polarization 
function only receives contributions from particle-hole pairs that are 
very close to the Fermi surface. On the other hand, the energy denominator 
diverges in this limit because the photon can excite particle-hole pairs 
with arbitrarily small energy. These two effects combine to give a finite 
contribution
\be
\label{pi00}
\Pi_{00}(q_0=0,\vec{q}\to 0) = 
e^2\int\!\frac{d^3p}{(2\pi)^3} \frac{\partial n_p}{\partial E_p}
 = e^2\frac{p_Fm}{2\pi^2},
\ee
which is proportional to the density of states on the Fermi surface. 
Equ.~(\ref{pi00}) implies that the static photon propagator in the 
limit $\vec{q}\to 0$ is modified according to $1/\vec{q}^{\,2} \to 
1/(\vec{q}^{\,2}+m_D^2)$, where 
\be 
\label{m_D}
m_D^2=e^2 \left( \frac{p_Fm}{2\pi^2}\right)
\ee
is called the Debye mass. The factor $N=(p_Fm)/(2\pi^2)$ is equal to 
the density of states on the Fermi surface. In a relativistic theory 
we find the same result as in equ.~(\ref{m_D}) with the density of 
states replaced by the correct relativistic expression $N=(p_FE_F)/
(2\pi^2)$. The Coulomb potential is modified as
\be 
V(r)= - e\frac{e^{-r/r_D}}{r},
\ee
where $r_D=1/m_D$ is called the Debye screening length. The physics 
of screening is very easy to understand. A test charge can polarize 
virtual particle-hole pairs that act to shield the charge. 

 In the same fashion we can study the response to an external vector 
potential $\vec{A}$. The coupling of a non-relativistic fermion to 
the vector potential is determined in the usual way by replacing 
$\vec{p}\to\vec{p}+e\vec{A}$. Since the kinetic energy operator 
is quadratic in the momentum we find a linear and a quadratic coupling 
of the vector potential. The one-loop diagrams that contribute to the 
polarization tensor are shown  in Fig.~\ref{fig_screen}. In the limit 
of small external momenta we find
\be
\label{pi_ij}
 \Pi_{ij}(q) = -e^2 m_D^2 \int\frac{d\Omega}{4\pi}\left\{
    v_i v_j \frac{\vec{v}\cdot\vec{q}}{q_0-\vec{v}\cdot\vec{q}}
  +\frac{1}{3}v^2\delta_{ij} \right\},
\ee
where $\vec{v}=\vec{p}/m$ is the Fermi velocity. In the limit $q_0\!=\!0$ 
the polarization tensor vanishes. There is no screening of static magnetic 
fields. For non-zero $q_0$ the trace of the polarization tensor is given by
\be
\label{pi_nr_ii}
\Pi_{ii}(q) = m_D^2\frac{vq_0}{2q}
  \log\left( \frac{q_0-vq}{q_0+vq}\right).
\ee
The result has an imaginary part for $vq>q_0$. This phenomenon is known 
as Landau damping. The physical mechanism is that the photon is loosing 
energy as it scatters of the electrons in the Fermi liquid, see 
\cite{Landau:kin} for a detailed discussion in the context of kinetic 
theory.

%%%%%%%%%%%%%%%%%%%%%%%%%%%%%%%%%%%%%%%%%%%%%%%%%%%%%%%%%%%%%%%%%%%%
\section{Superconductivity}
\label{sec_sc}
\subsection{BCS instability}
\label{sec_bcs}
%%%%%%%%%%%%%%%%%%%%%%%%%%%%%%%%%%%%%%%%%%%%%%%%%%%%%%%%%%%%%%%%%%%%

\begin{figure}[t]
\bc\includegraphics[width=9.0cm]{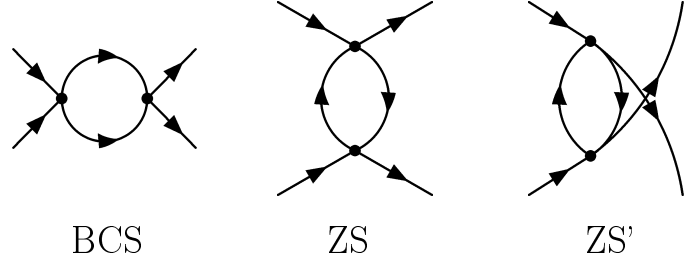}\ec
\caption{\label{fig_bcs}
Second order diagrams that contribute to particle-particle
scattering. The three diagrams are known as the ZS (zero sound),
ZS' and BCS (Bardeen-Cooper-Schrieffer) contribution.}
\end{figure}

 One of the most remarkable phenomena that take place in 
many body systems is superconductivity. Superconductivity 
is related to an instability of the Fermi surface in the 
presence of attractive interactions between fermions. Let 
us consider fermion-fermion scattering in the simple
model introduced in Sect.~\ref{sec_fl}. At leading 
order the scattering amplitude is given by
\be
\label{pp_0}
\Gamma_{\alpha\beta\gamma\delta}(p_1,p_2,p_3,p_4) = 
C_0 \left( \delta_{\alpha\gamma}\delta_{\beta\delta}
 - \delta_{\alpha\delta}\delta_{\beta\gamma} \right).
\ee
At next-to-leading order we find the corrections shown 
in Fig.~\ref{fig_bcs}. A detailed discussion of the role 
of these corrections can be found in 
\cite{Abrikosov:1963,Shankar:1993pf,Polchinski:1992ed}.
The BCS diagram is special, because in the case of a
spherical Fermi surface it can lead to an instability 
in weak coupling. The main point is that if the 
incoming momenta satisfy $\vec{p}_1\simeq -\vec{p}_2$
then there are no kinematic restrictions on the loop
momenta. As a consequence, all back-to-back pairs can
mix and there is an instability even in weak coupling. 

%%%%%%%%%%%%%%%%%%%%%%%%%%%%%%%%%%%%%%%%%%%%%%%%%%%%%%%%%%%%%%%%%%%%
%%%%%%%%%%%%%%%%%%%%%%%%%%%%%%%%%%%%%%%%%%%%%%%%%%%%%%%%%%%%%%%%%%%%

 For $\vec{p}_1= -\vec{p}_2$ and $E_1=E_2=E$ the BCS 
diagram is given by
\bea
\label{diag_bcs}
\Gamma_{\alpha\beta\gamma\delta} = 
C_0^2 \left( \delta_{\alpha\gamma}\delta_{\beta\delta}
 - \delta_{\alpha\delta}\delta_{\beta\gamma} \right)
\int \frac{d^4q}{(2\pi)^4} &&
 \frac{1}{E+q_0-\epsilon_q+i\delta{\rm sgn}(\epsilon_q)} 
  \nonumber \\
 &&  \frac{1}{E-q_0-\epsilon_q+i\delta{\rm sgn}(\epsilon_q)} .
\eea
As $E\to 0$ the loop integral develops an infrared divergence. This 
divergence comes from momenta near the Fermi surface and we can 
approximate $d^3q\simeq p_F^2dl$ with $l=|\vec{q}|-p_F$. The 
scattering amplitude is proportional to 
\be 
\label{cor_bcs}
\Gamma_{\alpha\beta\gamma\delta} =
\left( \delta_{\alpha\gamma}\delta_{\beta\delta}
 - \delta_{\alpha\delta}\delta_{\beta\gamma} \right)
\left\{
C_0 - C_0^2\left(\frac{p_Fm}{2\pi^2}\right)
 \log\left(\frac{E_0}{E}\right) \right\},
\ee
where $E_0$ is an ultraviolet cutoff. The logarithmic divergence
can also be seen by expanding equ.~(\ref{fpp_2}) around $s=0$ and
$\kappa=1$. The term in the curly brackets can be interpreted 
as an effective energy dependent coupling. The coupling constant satisfies 
the renormalization group equation \cite{Polchinski:1992ed,Shankar:1993pf}
\be 
\label{rge_bcs}
 E\frac{dC_0}{dE} = C_0^2 \left(\frac{p_Fm}{2\pi^2}\right),
\ee
with the solution
\be
\label{rge_sol}
C_0(E) =\frac{C_0(E_0)}{1+NC_0(E_0)\log(E_0/E)},
\ee
where $N=(p_Fm)/(2\pi^2)$ is the density of states. 
Equ.~(\ref{rge_sol}) shows that there are two possible
scenarios. If the initial coupling is repulsive, $C_0
(E_0)>0$, then the renormalization group evolution will 
drive the effective coupling to zero and the Fermi liquid 
is stable. If, on the other hand, the initial coupling is 
attractive, $C_0(E_0)<0$, then the effective coupling 
grows and reaches a Landau pole at 
\be 
\label{E_lp}
 E_{\it crit} \sim E_0 
    \exp\left(-\frac{1}{N|C_0(E_0)|}\right).
\ee
At the Landau pole the Fermi liquid description has to 
break down. The renormalization group equation does not
determine what happens at this point, but it seems 
natural to assume that the strong attractive interaction
will lead to the formation of a fermion pair condensate. 
The fermion condensate $\langle\epsilon^{\alpha\beta}
\psi_\alpha\psi_\beta\rangle$ signals the breakdown 
of the $U(1)$ symmetry and leads to a gap $\Delta$ in 
the single particle spectrum. 

%%%%%%%%%%%%%%%%%%%%%%%%%%%%%%%%%%%%%%%%%%%%%%%%%%%%%%%%%%%%%%%%%%%%
\begin{figure}[t]
\bc\includegraphics[width=11.0cm]{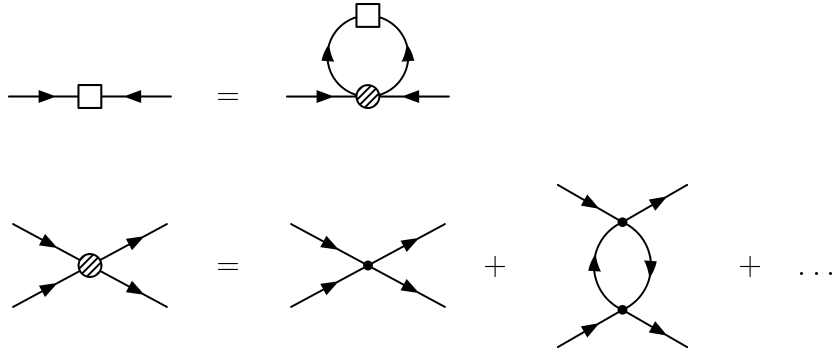}\ec
\caption{\label{fig_gap_bcs}
Gap equation for the superfluid gap in a theory with short 
range interactions.}
\end{figure}
%%%%%%%%%%%%%%%%%%%%%%%%%%%%%%%%%%%%%%%%%%%%%%%%%%%%%%%%%%%%%%%%%%%%

 The scale of the gap is determined by the position 
of the Landau pole, $\Delta\sim E_{\it crit}$. A more 
quantitative estimate of the gap can be obtained in the 
mean field approximation. In the path integral formulation 
the mean field approximation is most easily introduced
using the Hubbard-Stratonovich trick. For this purpose
we first rewrite the four-fermion interaction as
\be 
\label{4f_fierz}
\frac{C_0}{2}(\psi^\dagger\psi)^2  = 
\frac{C_0}{4} \left\{
 (\psi^\dagger\sigma_2\psi^\dagger)
 (\psi\sigma_2\psi) 
+(\psi^\dagger\sigma_2\vec{\sigma}\psi^\dagger)
 (\psi\vec{\sigma}\sigma_2\psi)\right\},
\ee 
where we have used the Fierz identity $2\delta^{\alpha
\beta}\delta^{\gamma\rho} = \delta^{\alpha\rho}
\delta^{\gamma\beta}+(\vec{\sigma})^{\alpha\rho}
(\vec{\sigma})^{\gamma\beta}$. Note that the second 
term in equ.~(\ref{4f_fierz}) vanishes because 
$(\sigma_2\vec{\sigma})$ is a symmetric matrix. We 
now introduce a factor of unity into the path integral
\be
1 = \frac{1}{Z_\Delta}\int D\Delta 
\exp\left(\frac{\Delta^*\Delta}{C_0}\right),
\ee
where we assume that $C_0<0$. We can eliminate the 
four-fermion term in the lagrangian by a shift in the 
integration variable $\Delta$. The action is now 
quadratic in the fermion fields, but it involves 
a Majorana mass term $\psi\sigma_2\Delta \psi+h.c$. The
Majorana mass terms can be handled using the Nambu-Gorkov 
method. We introduce the bispinor $\Psi=(\psi,\psi^\dagger
\sigma_2)$ and write the fermionic action as
\be
\label{s_ng}
{\cal S} = \frac{1}{2}\int\!\frac{d^4p}{(2\pi)^4}
 \Psi^\dagger
 \left(\begin{array}{cc}
     p_0-\epsilon_p  & \Delta \\
     \Delta^* & p_0+\epsilon_p
 \end{array}\right) \Psi.
\ee
Since the fermion action is quadratic we can integrate
the fermion out and obtain the effective lagrangian
\be
\label{s_ng_eff}
{\cal L}= \frac{1}{2}{\rm Tr}\left[\log\left(
 G_0^{-1}G\right)\right]+\frac{1}{C_0}|\Delta|^2,
\ee
where $G$ is the fermion propagator
\be
\label{ng_prop}
 G(p) = \frac{1}{p_0^2-\epsilon_p^2-|\Delta|^2}
 \left(\begin{array}{cc}
     p_0+\epsilon_p  & \Delta^* \\
     \Delta & p_0-\epsilon_p
 \end{array}\right).
\ee
The diagonal and off-diagonal components of $G(p)$ are sometimes 
referred to as normal and anomalous propagators. Note that we have not 
yet made any approximation. We have converted the fermionic path integral 
to a bosonic one, albeit with a very non-local action. The mean field 
approximation corresponds to evaluating the bosonic path integral using 
the saddle point method. Physically, this approximation means that the 
order parameter does not fluctuate. Formally, the mean field approximation 
can be justified in the large $N$ limit, where $N$ is the number of fermion 
fields. The saddle point equation for $\Delta$ gives the gap equation
\be
\Delta = |C_0|\int\!\frac{d^4p}{(2\pi)^4} 
 \frac{\Delta}{p_0^2-\epsilon^2_p-\Delta^2}.
\ee
Performing the $p_0$ integration we find
\be
\label{4f_gap}
1 = \frac{|C_0|}{2}\int\!\frac{d^3p}{(2\pi)^3} 
 \frac{1}{\sqrt{\epsilon^2_p+\Delta^2}}.
\ee
Since $\epsilon_p=E_p-\mu$ the integral in equ.~(\ref{4f_gap}) has 
an infrared divergence on the Fermi surface $|\vec{p}|\sim p_F$. As a 
result, the gap equation has a non-trivial solution even if the coupling 
is arbitrarily small. We can estimate the size of the gap as we did earlier
by writing $d^3p\simeq p_F^2dl$ and introducing a cutoff $\Lambda$ for the 
integral over $l$. We find $\Delta=2\Lambda\exp(-1/(N|C_0|))$. In order to 
obtain a more accurate result we compute the RHS of equ.~(\ref{4f_gap})
without the approximation $d^3q\simeq p_F^2dl$. We use dimensional 
regularization and
\be 
 \int_0^\infty dz\, \frac{z^\alpha}{\sqrt{(z-1)^2+x^2}} 
   = -\frac{\pi}{\sin(\pi\alpha)}  \left(1+x^2\right)^{\alpha/2}
 P_\alpha \left( - \frac{1}{\sqrt{1+x^2}}\right).
\ee
The dimensionally regularized gap equation is 
\cite{Papenbrock:1998wb,Marini:1998}
\be
\label{gap_dr}
 1 = \frac{\lambda\pi}{\sin(\pi\alpha)}
 \left(1+x^2\right)^{\alpha/2}
 P_\alpha \left( - \frac{1}{\sqrt{1+x^2}}\right) ,
\ee
where $2\lambda=C_0mp_F^{d-2}\Omega_d/(2\pi)^d$ is a dimensionless 
coupling constant, $\Omega_d$ is the surface area of the $d$-dimensional
unit ball and $x =\Delta/E_F$ is the dimensionless gap. $P_\alpha(z)$ is 
the Legendre function of order $\alpha$ and $\alpha=(d-2)/2$. Dimensional
regularization sets the power divergence in equ.~(\ref{4f_gap}) to zero. 
As a result, we can set $d=3$ and $C_0=4\pi a/m$ in equ.~(\ref{gap_dr}). 
If the gap is small, $x\ll 1$, equ.~(\ref{gap_dr}) can be solved using the 
asymptotic behavior of the Legendre function $P_\alpha(z)$ near the 
logarithmic singularity at $z=-1$,  
\be
P_\alpha(z) \simeq \frac{\sin(\alpha \pi)}{\pi}
 \left( \log\left(\frac{1+z}{2}\right) + 
 2\gamma+2\psi(\alpha+1) + \pi\cot(\alpha\pi) \right) .
\ee
We find
\be
\label{gap_bcs}
\Delta = \frac{8E_f}{e^2}\exp\left(-\frac{\pi}{2p_F|a|}\right).
\ee
The term in the exponent represents the leading term in an 
expansion in $p_F|a|$, see Fig.~\ref{fig_gap_bcs}. This means that 
in order to determine the pre-exponent in equ.~(\ref{gap_bcs})
we have to solve the gap equation at next-to-leading order. 
The contribution from the second diagram in Fig.~\ref{fig_gap_bcs}b
was first computed by Gorkov and Melik-Barkhudarov \cite{Gorkov:1961}.
The second order graph screens the leading order particle-particle
scattering amplitude and suppresses the $s$-wave gap by a factor
$(4e)^{1/3}\sim 2.2$ 

 For neutron matter the scattering length is large, $a=-18.8$ fm, and 
equ.~(\ref{gap_bcs}) is not very useful, except at very small density. 
At moderate density a rough estimate of the gap can be obtained by 
replacing $1/(p_Fa)$ with $\cot(\delta(k_F))$, where $\delta(k)$ is 
the $s$-wave phase shift. This estimate gives neutron gaps on the order 
of 1 MeV at nuclear matter density.

%%%%%%%%%%%%%%%%%%%%%%%%%%%%%%%%%%%%%%%%%%%%%%%%%%%%%%%%%%%%%%%%%%%%
\subsection{Superfluidity}
\label{sec_sfl}
%%%%%%%%%%%%%%%%%%%%%%%%%%%%%%%%%%%%%%%%%%%%%%%%%%%%%%%%%%%%%%%%%%%%

 Pairing leads to important physical effects. If the fermions are 
charged, pairing causes superconductivity. If the fermions are neutral, 
pairing leads to superfluidity. We first discuss superfluidity. The 
superfluid order parameter $\langle\psi\psi\rangle$ breaks the $U(1)$ 
symmetry and leads to the appearance of a Goldstone boson. The Goldstone 
boson field is defined as the phase of the order parameter
\be
  \langle\psi\psi\rangle  = |\langle\psi\psi\rangle| e^{2i\varphi}.
\ee
In the following we shall construct an effective lagrangian for the 
Goldstone field $\varphi$. The $U(1)$ symmetry implies that the lagrangian 
can only depend on derivatives of $\varphi$. The simplest possibility
is 
\be 
\label{sfl_gb_eft}
{\cal L} = f^2 \left( (\partial_0\varphi)^2
    - v^2(\partial_i\varphi)^2 + \ldots \right) ,
\ee
where $vf$ is the coupling of $\partial_i\varphi$ to the $U(1)$ 
current and $v$ is the Goldstone boson velocity. This Lagrangian 
correctly describes the propagation of Goldstone modes and the 
coupling to external currents, but it does not respect Galilean 
invariance, and it does not describe the interaction between 
Goldstone modes \cite{Greiter:1989qb,Son:2002zn,Son:2005rv}. 
Under Galilean transformations the fermion field transforms as
\be
\label{Gal_trafo}
 \psi(t,\vec{x}) \to \psi'(t,\vec{x}) = 
   e^{im\vec{v}\cdot \vec{x}} \psi(t,\vec{x}-\vec{v} t) .
\ee
This implies that $\varphi$ transforms as $\varphi(t,\vec{x})\to 
\varphi(t,\vec{x}-\vec{v} t)+m\vec{v}\cdot \vec{x}$. We also observe 
that the chemical potential enters the microscopic theory like the 
time component of a $U(1)$ gauge field. We can impose the constraints 
of Galilei invariance and $U(1)$ symmetry by constructing
an effective lagrangian that only depends on the variable
\be
X = \mu - \partial_0\varphi - \frac{(\partial_i\varphi)^2}{2m}.
\ee
In the following it will be useful to consider a low energy expansion 
in which $\partial_0\varphi$, $\partial_i\varphi$ are $O(1)$ but higher 
derivatives $\partial_i\partial_j\varphi$, etc.~are suppressed. In 
this case the leading order lagrangian contains arbitrary powers of 
$X$, but terms with derivatives of $X$ are suppressed. The functional
form of ${\cal L}(X)$ can be determined using the following simple
argument. For constant fields $\varphi=const$ the lagrangian ${\cal L}
(X)={\cal L}(\mu)$ is equal to minus the thermodynamic potential $\Omega$. 
Since $\Omega=-P$, where $P$ is the pressure, we conclude that ${\cal L}
(X)=P(X)$.

 As an example consider superfluidity in a weakly coupled Fermi 
gas. At leading order the equation of state is that of a free 
Fermi gas, $P = m^{3/2}(2\mu)^{5/2}/(15\pi^2)$.  The effective
lagrangian is given by
\be
\label{l_sfl_pert}
  {\cal L} = \frac{2^{5/2} m^{3/2}}{15\pi^2} \left(\mu -\partial_0\varphi
      -\frac{(\partial_i\varphi)^2}{2m} \right)^{5/2}.
\ee
We can determine the Goldstone boson propagator as well as
Goldstone boson interactions by expanding this result in powers 
of $\partial_0\varphi$ and $\partial_i\varphi$. There are some 
predictions that are independent of the equation of state. 
Consider the effective theory at second order in $(\partial\varphi)$,
\be
  {\cal L} = P(\mu) - n\partial_0\varphi 
   + \frac{1}{2}\frac{\partial n}{\partial\mu}(\partial_0\varphi)^2
   - \frac{n}{2m}(\partial_i\varphi)^2 + \ldots, 
\ee
where we have used $n=(\partial P)/(\partial\mu)$. The Goldstone boson
velocity is given by 
\be
  v^2 = \frac{n}{m} \frac{\partial\mu}{\partial n} 
      = \frac{\partial P}{\partial\rho} \,.
\ee
where $\rho=nm$ denotes the mass density.  We observe that the Goldstone 
boson velocity is given by the same formula as the speed of sound in a 
normal fluid. In a weakly interacting Fermi gas $v^2=v_F^2/3$.

 It is also instructive to study the relation to fluid dynamics in 
more detail. The equation of motion for the field $\varphi$ is given by
\be
\label{hyd_cont}
  \partial_0 \bar{n} + \frac{1}{m} 
    \vec\nabla \left(\bar{n}\vec\nabla \varphi\right) =0,
\ee
where we have defined $\bar{n}=P'(X)$. Equ.~(\ref{hyd_cont}) is the 
continuity equation for the current $j_\mu=\bar{n}(1,\vec{v}_s)$ where we
have identified the fluid velocity
\be
\label{def_vs}
  \vec{v}_s = \frac{\vec{\nabla}\varphi}{m}\,.
\ee
In the hydrodynamic description the independent variables are $\bar{n}$ and
$\vec{v}_s$. We can derive a second equation by using the identity
$dP=nd\mu$. We get 
\be
 \label{euler}
  \partial_0 \vec{v}_s +\frac{1}{2} \vec\nabla v_s^2 = 
  -\frac{1}{m}\vec{\nabla}\mu.
\ee
This is the Euler equation for non-viscous, irrotational fluid. The
fact that the flow is irrotational follows from the definition of 
the velocity as the gradient of $\varphi$. We conclude that the low
energy effective lagrangian is equivalent to superfluid hydrodynamics.

%%%%%%%%%%%%%%%%%%%%%%%%%%%%%%%%%%%%%%%%%%%%%%%%%%%%%%%%%%%%%%%%%%%%
\subsection{Landau-Ginzburg theory}
\label{sec_lg}
%%%%%%%%%%%%%%%%%%%%%%%%%%%%%%%%%%%%%%%%%%%%%%%%%%%%%%%%%%%%%%%%%%%%

 In this section we shall study the properties of a superconductor 
in more detail. Superconductors are characterized by the fact that 
the $U(1)$ symmetry is gauged. The order parameter $\Phi=\langle 
\epsilon^{\alpha\beta}\psi_\alpha\psi_\beta\rangle$ breaks $U(1)$ 
invariance. Consider a gauge transformation 
\be 
A_\mu\to A_\mu +\partial_\mu\Lambda .
\ee
The order parameter transforms as
\be 
\Phi \to \exp(2ie\Lambda)\Phi.
\ee
The breaking of gauge invariance is responsible for most of the 
unusual properties of superconductors \cite{Anderson:1984,Weinberg:1995}.
This can be seen by constructing the low energy effective action 
of a superconductor. For this purpose we write the order parameter
in terms of its modulus and phase
\be 
\Phi(x) = \exp(2ie\phi(x)) \tilde\Phi(x).
\ee
The field $\phi$ corresponds to the Goldstone mode. Under a gauge
transformation $\phi(x)\to\phi(x)+\Lambda(x)$. Gauge invariance restricts 
the form of the effective Lagrange function as
\be 
\label{L_sc}
 L = -\frac{1}{4}\int d^3x\, F_{\mu\nu}F_{\mu\nu}
 + L_s (A_\mu-\partial_\mu\phi).
\ee
There is a large amount of information we can extract even without 
knowing the explicit form of $L_s$. Stability implies that $A_\mu=
\partial_\mu\phi$ corresponds to a minimum of the energy. This means 
that up to boundary effects the gauge potential is a total divergence 
and that the magnetic field has to vanish. This phenomenon is known as 
the Meissner effect. 

 Equ.~(\ref{L_sc}) also implies that a superconductor has zero resistance. 
The equations of motion relate the time dependence of the Goldstone boson 
field to the potential, 
\be
\label{phidot}
\dot\phi(x)=-V(x).
\ee 
The electric current is related to the gradient of the Goldstone boson 
field. Equ.~(\ref{phidot}) shows that the time dependence of the current 
is proportional to the gradient of the potential. In order to obtain a 
static current the gradient of the potential has to vanish throughout 
the sample, and the resistance is zero. 

 In order to study the properties of a superconductor in more detail we 
have to specify $L_s$. For this purpose we assume that the system is 
time-independent, that the spatial gradients are small, and that the 
order parameter is small. In this case we can write
\be 
\label{l_lg}
L_s = \int d^3x\, \left\{
-\frac{1}{2}\left|\left(\vec\nabla-2ie\vec{A}\right)\Phi\right|^2
 +\frac{1}{2}m^2_H\left(\Phi^*\Phi\right)^2
 -\frac{1}{4}g\left(\Phi^*\Phi\right)^4 + \ldots \right\},
\ee
where $m_H$ and $g$ are unknown parameters that depend on the temperature. 
Equ.~(\ref{l_lg}) is known as the Landau-Ginzburg effective action. 
Strictly speaking, the assumption that the order parameter is small can 
only be justified in the vicinity of a second order phase transition. 
Nevertheless, the Landau-Ginzburg description is instructive even in the 
regime where $t=(T-T_c)/T_c$ is not small. It is useful to decompose 
$\Phi=\rho\exp(2ie\phi)$. For constant fields the effective potential, 
\be
\label{v_lg}
V(\rho)=-\frac{1}{2}m_H^2\rho^2 +\frac{1}{4}g\rho^4 ,
\ee
is independent of $\phi$. The minimum is at $\rho_0^2=m_H^2/g$ and the 
energy density at the minimum is given by ${\cal E}= -m_H^4/(4g)$. This 
shows that the two parameters $m_H$ and $g$ can be related to the 
expectation value of $\Phi$ and the condensation energy. We also observe 
that the phase transition is characterized by $m_H(T_c)=0$. 

 In terms of $\phi$ and $\rho$ the Landau-Ginzburg action is given by
\be 
L_s = \int d^3x\, \left\{
-2e^2\rho^2 \left(\vec\nabla\phi-\vec{A}\right)^2
 +\frac{1}{2}m_H^2\rho^2 -\frac{1}{4}g\rho^4
 -\frac{1}{2}\left(\vec\nabla\rho\right)^2
\right\}.
\ee
The equations of motion for $\vec{A}$ and $\rho$ are given by
\bea 
\label{b_lg}
\vec\nabla\times \vec{B} &=& 
 4e^2\rho^2 \left(\vec\nabla\phi -\vec{A}\right), \\
\label{rho_lg}
 \vec\nabla^2 \rho &=& 
 -m_H^2\rho^2 + g\rho^3 + 4e^2 \rho 
 \left( \vec\nabla\phi-\vec{A}\right)^2 .
\eea
Equ.~(\ref{b_lg}) implies that $\nabla^2\vec{B} = -4e^2\rho^2\vec{B}$. 
This means that an external magnetic field $\vec{B}$ decays over a 
characteristic distance $\lambda=1/(2e\rho)$. Equ.~(\ref{rho_lg}) gives 
$\nabla^2\rho = -m_H^2\rho+\ldots$. As a consequence, variations in the 
order parameter relax over a length scale given by $\xi=1/m_H$. The two 
parameters $\lambda$ and $\xi$ are known as the penetration depth and the 
coherence length. 

 The relative size of $\lambda$ and $\xi$ has important consequences for 
the properties of superconductors. In a type II superconductor $\xi<\lambda$. 
In this case magnetic flux can penetrate the system in the form of vortex 
lines. At the core of a vortex the order parameter vanishes, $\rho=0$. In 
a type II material the core is much smaller than the region over which the 
magnetic field goes to zero. The magnetic flux is given by
\be
\int_A\vec{B}\cdot d\vec{S} =
\oint_{\partial A} \vec{A}\cdot d\vec{l} = 
\oint_{\partial A} \vec{\nabla}\phi \cdot d\vec{l} =
\frac{n\pi\hbar}{e} ,
\ee
and quantized in units of $\pi\hbar/e$. In a type II superconductor 
magnetic vortices repel each other and form a regular lattice known as 
the Abrikosov lattice. In a type I material, on the other hand, vortices 
are not stable and magnetic fields can only penetrate the sample if 
superconductivity is destroyed. 

%%%%%%%%%%%%%%%%%%%%%%%%%%%%%%%%%%%%%%%%%%%%%%%%%%%%%%%%%%%%%%%%%%%%
\subsection{Microscopic calculation of the screening mass}
\label{sec_mei}
%%%%%%%%%%%%%%%%%%%%%%%%%%%%%%%%%%%%%%%%%%%%%%%%%%%%%%%%%%%%%%%%%%%%

%%%%%%%%%%%%%%%%%%%%%%%%%%%%%%%%%%%%%%%%%%%%%%%%%%%%%%%%%%%%%%%%%%%%
\begin{figure}[t]
\bc\includegraphics[width=8.0cm]{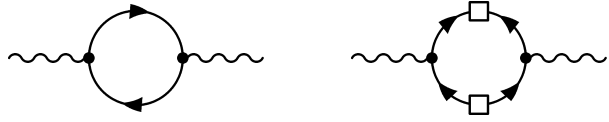}\ec
\caption{\label{fig_mei}
Leading order Feynman diagrams that contribute to the 
photon polarization function in a superconducting Fermi gas.
The figure does not show the tadpole diagram.}
\end{figure}
%%%%%%%%%%%%%%%%%%%%%%%%%%%%%%%%%%%%%%%%%%%%%%%%%%%%%%%%%%%%%%%%%%%%

 In this section we shall study screening of gauge fields in a 
superconductor from a more microscopic point of view. The calculation
is analogous to the one discussed in Sect.~\ref{sec_screen}. The
difference is that the propagators contain the gap, and that there 
is an extra trace over Nambu-Gorkov indices, see Fig.~\ref{fig_mei}. 
The polarization functions contains normal contributions proportional 
to $G_{11}G_{11}$ and $G_{22}G_{22}$ as well as anomalous terms 
proportional to $G_{12}G_{21}$, where $G_{ij}$ is the Nambu-Gorkov 
propagator give in equ.~(\ref{ng_prop}). The sum of the normal and 
anomalous diagrams is given by
\be
\label{pi00_ng}
\Pi_{00}(q\!=\!0) =-ie^2
  \int\!\frac{d^4p}{(2\pi)^4} 
   \left\{
  \frac{p_0^2+\epsilon_p^2}{(p_0^2-\epsilon_p^2-\Delta^2)^2} 
  -\frac{\Delta^2}{ (p_0^2-\epsilon_p^2-\Delta^2)^2} \right\}.
\ee
The integral over $p_0$ can be done by contour integration. The two 
terms in equ.~(\ref{pi00_ng}) give equal contributions. We find
\be
\Pi_{00}(q\!=\!0)  = 
  e^2 \int\! \frac{d^3p}{(2\pi)^3}
           \frac{\Delta^2}{(\epsilon_p^2+\Delta^2)^{3/2}} .
\ee
This integral is dominated by very small energies $|\epsilon_p|=
|E_p-\mu|\sim\Delta$ and we can approximate $\epsilon_p =v_F (p-p_F)$.
We find
\be
\Pi_{00}(q\!=\!0) =  e^2\;\frac{p_Fm}{2\pi^2},
\ee
which is identical to the result in the normal phase. There are a 
number of subtleties that are worth commenting on. First we note that 
the polarization function in the superfluid phase is analytic in 
the external momenta and we can set $q_0=\vec{q}=0$ from the beginning. 
We also note that the normal contribution is formally ultraviolet 
divergent. The correct prescription to deal with this divergence 
is to perform the $p_0$ integral first \cite{Abrikosov:1963}. Finally
we observe that while the screening masses in the normal and superfluid
phase are the same, only half of the result in the superfluid phase
is contributed by the normal term. 

 The calculation of the electric polarization function is easily 
generalized to the magnetic case. There are three diagrams. The
first is the tadpole contribution discussed in Sect.~\ref{sec_screen}.
This contribution is proportional to the total density and is the 
same in the normal and superfluid phase. The normal and anomalous
one-loop diagrams are similar to the electric case, but the coupling
$e^2$ is replaced by $e^2v_iv_j$ in the normal contribution and 
$e^2v_i(-v_j)$ in the anomalous term. As a result the two terms 
cancel and the polarization function is given by the tadpole term
\be
\Pi_{ij}(q\!=\!0) = -e^2v_F^2\delta_{ij}\,\frac{p_Fm}{6\pi^2}.
\ee
We find that there is a non-zero magnetic screening mass in the 
superfluid phase, and that the Meissner mass is controlled not by 
the gap, but by the density of states on the Fermi surface. This 
does not contradict the fact that the magnetic screening mass goes
to zero as $\Delta\to 0$. We find that the photon mass term
has the structure $m_D^2(A_0^2-v_F^2\vec{A}^2/3)$. This result can
also be obtained by gauging the effective Lagrangian for the 
Goldstone boson, equ.~(\ref{sfl_gb_eft}), together with the result 
$v^2=v_F^2/3$ for the speed of sound in a weakly interacting
Fermi gas.

%%%%%%%%%%%%%%%%%%%%%%%%%%%%%%%%%%%%%%%%%%%%%%%%%%%%%%%%%%%%%%%%%%%%
\section{Strongly interacting fermions}
\label{sec_uni}
%%%%%%%%%%%%%%%%%%%%%%%%%%%%%%%%%%%%%%%%%%%%%%%%%%%%%%%%%%%%%%%%%%%%

 Up to this point we have concentrated on weakly coupled many body 
systems. In this section we shall consider a cold, dilute gas of 
fermionic atoms in which the scattering length $a$ of the atoms can 
be changed continuously. This system can be realized experimentally 
using Feshbach resonances, see \cite{Regal:2005} for a review. A 
small negative scattering length corresponds to a weak attractive 
interaction between the atoms. This case is known as the BCS limit. 
As the strength of the interaction increases the scattering length 
becomes larger. It diverges at the point where a bound state is formed. 
The point $a=\infty$ is called the unitarity limit, since the scattering 
cross section saturates the $s$-wave unitarity bound $\sigma=4\pi/k^2$. 
On the other side of the resonance the scattering length is positive. 
In the BEC limit the interaction is strongly attractive and the fermions 
form deeply bound molecules.

%%%%%%%%%%%%%%%%%%%%%%%%%%%%%%%%%%%%%%%%%%%%%%%%%%%%%%%%%%%%%%%%%%%%
\begin{figure}[t]
\begin{minipage}{0.49\hsize}
\bc\includegraphics[width=0.95\hsize]{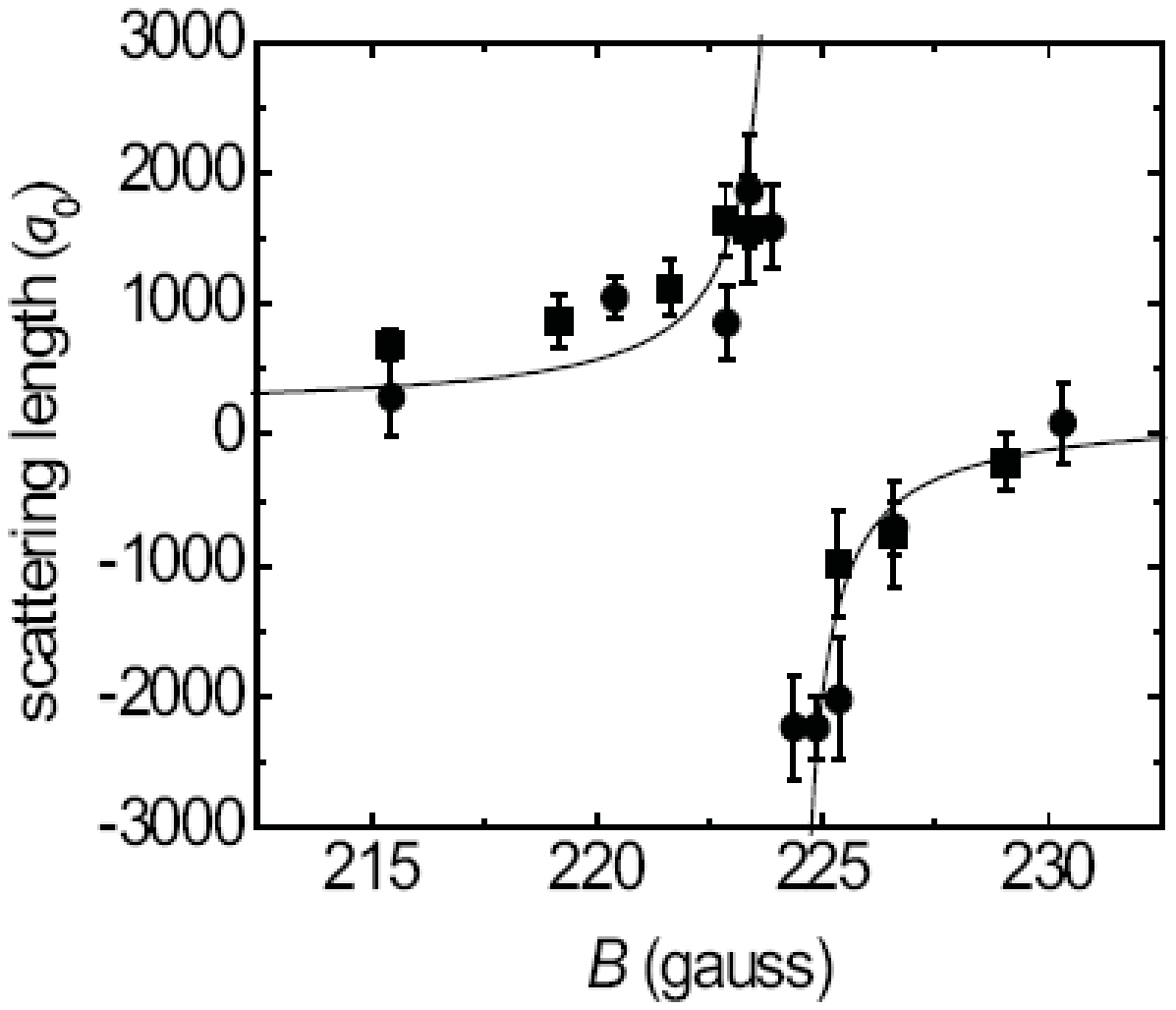}\ec
\end{minipage}\begin{minipage}{0.49\hsize}
\includegraphics[width=0.95\hsize]{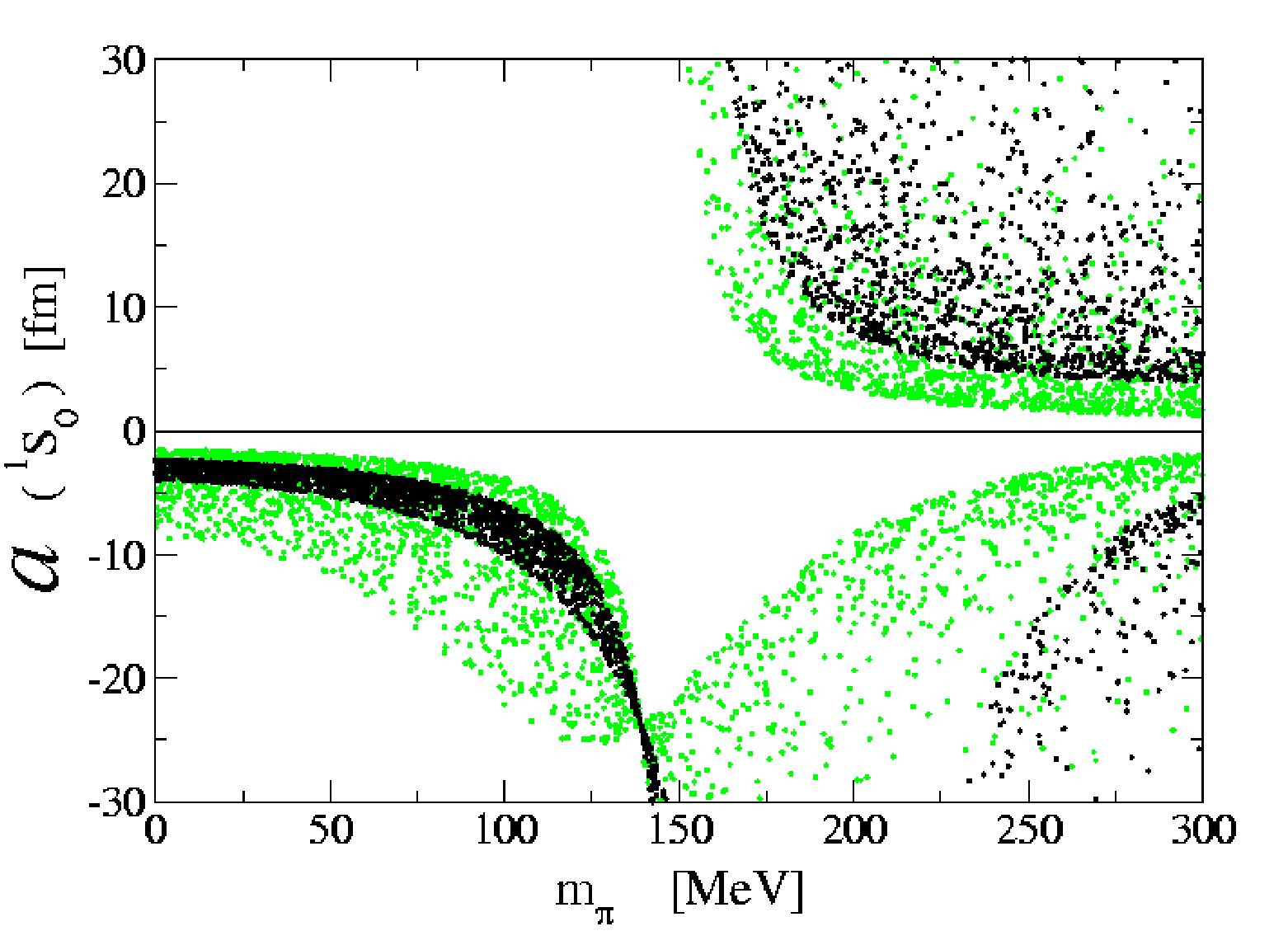}
\end{minipage}
\caption{\label{fig_fesh}
The left panel shows the scattering length of $^{40}$K Atoms
as a function of the magnetic field near a Feshbach resonance, from 
Regal (2005). The right panel shows the nucleon-nucleon scattering 
length in the $^1S_0$ channel as a function of the pion mass. The 
scatter plot indicates the uncertainty due to higher order terms
in the chiral effective lagrangian. Figure from Beane \& Savage
(2003).}
\end{figure}
%%%%%%%%%%%%%%%%%%%%%%%%%%%%%%%%%%%%%%%%%%%%%%%%%%%%%%%%%%%%%%%%%%%%

 A dilute gas of fermions in the unitarity limit is a strongly coupled
quantum liquid that exhibits many interesting properties. One 
interesting feature is universality. We are interested in the limit 
$(k_Fa)\to\infty$ and $(k_Fr)\to 0$, where $k_F$ is the Fermi momentum, 
$a$ is the scattering length and $r$ is the effective range. From 
dimensional analysis it is clear that the energy per particle at zero
temperature has to be proportional to energy per particle of a free Fermi 
gas at the same density
\be
\frac{E}{A} = \xi \Big(\frac{E}{A}\Big)_0 = \xi 
\frac{3}{5}\Big(\frac{k_F^2}{2m}\Big).
\ee
The constant $\xi$ is universal, i.~e.~ independent of the details of 
the system. Similar universal constants govern the magnitude of the 
gap in units of the Fermi energy and the equation of state at finite
temperature. 

 Universal behavior in the unitarity limit is relevant to the physics 
of dilute neutron matter. The neutron-neutron scattering length is 
$a_{nn}=-18$ fm and the effective range is $r_{nn}=2.8$ fm. This means 
that there is a range of densities for which the inter-particle spacing 
is large compared to the effective range but small compared to the 
scattering length. It is interesting to note that the neutron scattering 
length depends on the quark masses in a way that is very similar to the 
dependence of atomic scattering lengths on the magnetic field near a 
Feshbach resonance \cite{Beane:2002xf}, see Fig.~\ref{fig_fesh}.

%%%%%%%%%%%%%%%%%%%%%%%%%%%%%%%%%%%%%%%%%%%%%%%%%%%%%%%%%%%%%%%%%%%%
\subsection{Numerical Calculations}
\label{sec_unit_latt}
%%%%%%%%%%%%%%%%%%%%%%%%%%%%%%%%%%%%%%%%%%%%%%%%%%%%%%%%%%%%%%%%%%%%

%%%%%%%%%%%%%%%%%%%%%%%%%%%%%%%%%%%%%%%%%%%%%%%%%%%%%%%%%%%%%%%%%%%%
\begin{figure}[t]
\bc\includegraphics[angle=-90,width=0.75\hsize]{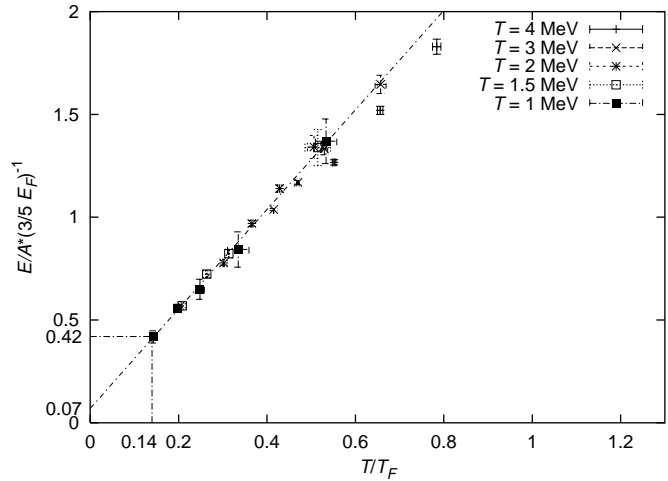}\ec
\caption{\label{fig_latt}
Lattice results for the energy per particle of a dilute Fermi gas 
from Lee \& Sch\"afer (2005). We show the energy per particle 
in units of $3E_F/5$ as a function of temperature in units of 
$T_F$.}
\end{figure}
%%%%%%%%%%%%%%%%%%%%%%%%%%%%%%%%%%%%%%%%%%%%%%%%%%%%%%%%%%%%%%%%%%%%

 The calculation of the dimensionless quantity $\xi$ is a non-perturbative 
problem. In this section we shall tackle this problem using a combination 
of effective field theory and lattice field theory methods. We will study
an analytical approach in the next section. We first observe that in the 
low density limit the details of the interaction are not important. The 
physics of the unitarity limit is captured by an effective lagrangian of 
point-like fermions interacting via a short-range interaction. The 
lagrangian is 
\be 
{\cal L} = \psi^\dagger \left( i\partial_0 
      + \frac{\vec\nabla^2}{2m} \right) \psi 
      - \frac{C_0}{2} \left(\psi^\dagger \psi\right)^2 ,
\ee
as in Equ.~(\ref{l_4f}). The usual strategy for dealing with the 
four-fermion interaction is to use a Hubbard-Stratonovich transformation
as in Sect.~\ref{sec_bcs}. The partition function can be written 
as \cite{Lee:2004qd}
\be
Z = \int DsDcDc^{\ast} \exp\left[-S\right]  ,
\ee
where $s$ is the Hubbard-Stratonovich field and $c$ is a Grassmann field. 
$S$ is a discretized euclidean action
\bea
S  &=& 
 \sum_{\vec{n},i}\left[  e^{-\hat\mu\alpha_{t}}c_{i}^{\ast}
   (\vec{n})c_{i} (\vec{n}+\hat{0})-e^{\sqrt{-C_0\alpha_{t}}
  s(\vec{n})+\frac{C_0\alpha_{t}}{2}}(1-6h)c_{i}^{\ast}
   (\vec{n})c_{i} (\vec{n})\right] \nonumber\\
& & \hspace{0.3cm}\mbox{} 
   -h\sum_{\vec{n},l_{s},i}\left[  
   c_{i}^{\ast}(\vec{n})c_{i}(\vec{n}
   +\hat{l}_{s})+c_{i}^{\ast}(\vec{n})c_{i}
  (\vec{n}-\hat{l}_{s})\right]  +\frac{1}{2}\sum_{\vec{n}}s^{2}(\vec{n}).
\eea
Here $i$ labels spin and $\vec{n}$ labels lattice sites. Spatial and
temporal unit vectors are denoted by $\hat{l}_s$ and $\hat{0}$, 
respectively. The temporal and spatial lattice spacings are $b_\tau$
and $b$. The dimensionless chemical potential is given by $\hat{\mu}
=\mu b_\tau$. We define $\alpha_t$ as the ratio of the temporal and 
spatial lattice spacings and $h=\alpha_t/(2\hat{m})$. Note that for 
$C_0<0$ the action is real and standard Monte Carlo simulations are 
possible. 

 The four-fermion coupling is fixed by computing the sum of all 
particle-particle bubbles as in Sect.~\ref{sec_nr_eft} but with the 
elementary loop function regularized on the lattice. Schematically, 
\be 
\frac{m}{4\pi a} = \frac{1}{C_0} 
   + \frac{1}{2} \sum_{\vec{p}}\frac{1}{E_{\vec{p}}} ,
\ee
where the sum runs over discrete momenta on the lattice and $E_{\vec{p}}$ 
is the lattice dispersion relation. A detailed discussion of the 
lattice regularized scattering amplitude can be found in 
\cite{Chen:2003vy,Beane:2003da,Lee:2004qd}. For a given scattering 
length $a$ the four-fermion coupling is a function of the lattice 
spacing. The continuum limit correspond to taking the temporal 
and spatial lattice spacings $b_\tau$, $b$ to zero
\be 
 b_\tau\mu\to 0, \hspace{1cm} bn^{1/3}\to 0 ,
\ee
keeping $an^{1/3}$ fixed. Here, $\mu$ is the chemical potential and
$n$ is the density. Numerical results in the unitarity limit are shown 
in Fig.~\ref{fig_latt}. From these simulations we concluded that $\xi=
(0.09-0.42)$. Lee performed canonical simulations at $T=0$ and obtained 
\cite{Lee:2005fk} $\xi=0.25$. Green Function Monte Carlo calculations 
give \cite{Carlson:2003wm} $\xi=0.44$, and finite temperature lattice 
simulations have been extrapolated to $T=0$ to yield similar results 
\cite{Bulgac:2005pj,Burovski:2006}.

%%%%%%%%%%%%%%%%%%%%%%%%%%%%%%%%%%%%%%%%%%%%%%%%%%%%%%%%%%%%%%%%%%%%
\subsection{Epsilon Expansion}
\label{sec_eps}
%%%%%%%%%%%%%%%%%%%%%%%%%%%%%%%%%%%%%%%%%%%%%%%%%%%%%%%%%%%%%%%%%%%%

 It is also desirable to find a systematic analytical approach to 
the dilute Fermi liquid in the unitarity limit. Various possibilities
have been considered, such as an expansion in the number of fermion
species \cite{Furnstahl:2002gt,Nikolic:2006} or the number of spatial 
dimensions \cite{Steele:2000qt,Schafer:2005kg}. Nussinov \& Nussinov 
observed that the fermion many body system in the unitarity limit 
reduces to a free Fermi gas near $d=2$ spatial dimensions, and to a 
free Bose gas near $d=4$ \cite{Nussinov:2004}. Their argument was based
on the behavior of the two-body wave function as the binding energy 
goes to zero. For $d=2$ it is well known that the limit of zero 
binding energy corresponds to an arbitrarily weak potential. In $d=4$ 
the two-body wave function at $a=\infty$ has a $1/r^2$ behavior and 
the normalization is concentrated near the origin. This suggests 
the many body system is equivalent to a gas of non-interacting bosons.

 A systematic expansion based on the observation of Nussinov \& Nussinov
was studied by Nishida and Son \cite{Nishida:2006br,Nishida:2006eu}. In
this section we shall explain their approach. We begin by restating the 
argument of Nussinov \& Nussinov in the effective field theory language. 
In dimensional regularization $a\to\infty$ corresponds to $C_0\to\infty$. 
The fermion-fermion scattering amplitude (see equ.~\ref{nn_sum}) is 
given by
\be 
{\cal A}(p_0,\vec{p}) =  \left(\frac{4\pi}{m}\right)^{d/2}
 \left[\Gamma\left(1-\frac{d}{2}\right)\right]^{-1} 
 \frac{i}{\left(-p_0+E_p/2-i\delta\right)^{\frac{d}{2}-1}}\; ,
\ee
where $\delta\to 0+$. As a function of $d$ the Gamma function has poles 
at $d=2,4,\ldots$ and the scattering amplitude vanishes at these points. 
Near $d=2$ the scattering amplitude is energy and momentum independent.
For $d=4-\epsilon$ we find
\be
\label{A_4-eps}
{\cal A}(p_0,\vec{p}) =  \frac{8\pi^2\epsilon}{m^2}
 \frac{i}{p_0-E_p/2+i\delta} + O(\epsilon^2) \, .
\ee
We observe that at leading order in $\epsilon$ the scattering amplitude 
looks like the propagator of a boson with mass $2m$. The boson-fermion
coupling is $g^2=(8\pi^2\epsilon)/m^2$ and vanishes as $\epsilon\to 0$. 
This suggests that we can set up a perturbative expansion involving 
fermions of mass $m$ weakly coupled to bosons of mass $2m$. In the 
unitarity limit the Hubbard-Stratonovich transformed lagrangian reads
\be
{\cal L}= \Psi^\dagger\left[
     i\partial_0+\sigma_3\frac{\vec\nabla^2}{2m}\right]\Psi
  + \mu\Psi^\dagger\sigma_3\Psi
  +\left(\Psi^\dagger\sigma_+\Psi\phi + h.c. \right)\ ,\nonumber
\ee
where $\Psi=(\psi_\uparrow,\psi_\downarrow^\dagger)^T$ is a two-component 
Nambu-Gorkov field, $\sigma_i$ are Pauli matrices acting in the Nambu-Gorkov 
space and $\sigma_\pm=(\sigma_1\pm i\sigma_2)/2$. In the superfluid phase
$\phi$ acquires an expectation value. We write 
\be
 \phi = \phi_0 + g\varphi, \hspace{1cm}
   g  =\frac{\sqrt{8\pi^2\epsilon}}{m}
       \left(\frac{m\phi_0}{2\pi}\right)^{\epsilon/4} ,
\ee
where $\phi_0=\langle\phi\rangle$. The scale $M^2=m\phi_0/(2\pi)$ was
introduced in order to have a correctly normalized boson field. The scale 
parameter is arbitrary, but this particular choice simplifies some of the 
loop integrals. In order to get a well defined perturbative expansion we 
add and subtract a kinetic term for the boson field to the lagrangian. We
include the kinetic term in the free part of the lagrangian
\be
{\cal L}_0 = \Psi^\dagger\left[i\partial_0+\sigma_3\frac{\vec\nabla^2}{2m}
     + \phi_0(\sigma_{+} +\sigma_{-})\right]\Psi
     + \varphi^\dagger\left(i\partial_0
        + \frac{\vec\nabla^2}{4m}\right)\varphi\, .
\ee
The interacting part is 
\be
{\cal L}_I = g\left(\Psi^\dagger\sigma_+\Psi\varphi + h.c\right)
     + \mu\Psi^\dagger\sigma_3\Psi 
     - \varphi^\dagger\left(i\partial_0
        + \frac{\vec\nabla^2}{4m}\right)\varphi\, .
\ee
Note that the interacting part generates self energy corrections to 
the boson propagator which, by virtue of equ.~(\ref{A_4-eps}), cancel 
against the kinetic term of boson field. We have also included the 
chemical potential term in ${\cal L}_I$. This is motivated by the fact 
that near $d=4$ the system reduces to a non-interacting Bose gas and
$\mu\to 0$. We will count $\mu$ as a quantity of $O(\epsilon)$. 

 The Feynman rules are quite simple. The fermion and boson propagators 
are
\bea
\label{eps_prop}
G(p_0,\vec p) &=& \frac{i}{p_0^2-E_{\vec p}^2-\phi_0^2}
\left[\begin{array}{cc}
    p_0+E_{\vec p} &  -\phi_0\\
    -\phi_0        & p_0-E_{\vec p}
\end{array}\right]  \ ,\\
D(p_0, \vec p) &=& \frac{i}{p_0-E_{\vec p}/2}\ , 
\eea 
and the fermion-boson vertices are $ig\sigma^\pm$. Insertions of the 
chemical potential are $i\mu\sigma_3$. Both $g^2$ and $\mu$ are 
corrections of order $\epsilon$. In order to verify that the $\epsilon$
expansion is well defined we have to check that higher order diagrams
do not generate powers of $1/\epsilon$. Studying the superficial degree
of divergence of diagrams involving the propagators given in 
equ.~(\ref{eps_prop}) one can show that there is only a finite number 
of one-loop diagrams that generate $1/\epsilon$ terms. 

%%%%%%%%%%%%%%%%%%%%%%%%%%%%%%%%%%%%%%%%%%%%%%%%%%%%%%%%%%%%%%%%%%%%
\begin{figure}[t]
\bc\includegraphics[width=0.75\hsize]{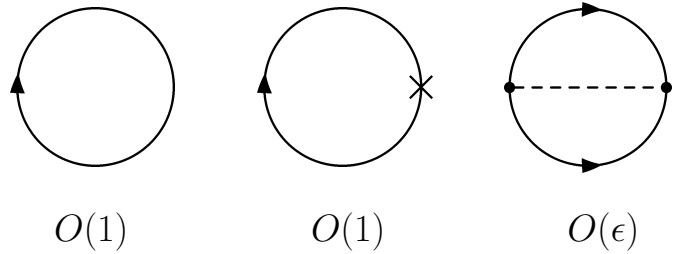}\ec
\caption{\label{fig_veff_eps}
Leading order contributions to the effective potential in 
the $\epsilon$ expansion. Solid lines are fermion propagators, 
dashed lines are boson propagators, and the cross is an insertion
of the chemical potential.}
\end{figure}
%%%%%%%%%%%%%%%%%%%%%%%%%%%%%%%%%%%%%%%%%%%%%%%%%%%%%%%%%%%%%%%%%%%%

 The leading order diagrams that contribute to the effective potential 
are shown in Fig.~\ref{fig_veff_eps}. The first diagram is the free
fermion loop which is $O(1)$. The second diagram is the $\mu$ insertion
which is $O(1)$ because the loop diagram is divergent in $d=4$. The
two-loop diagram is $O(\epsilon)$ because of the factor of $g^2$ from 
the vertices. The free fermion loop diagram is
\be
 V_0 =i\int\frac{dp_0}{2\pi}\int\frac{d^dp}{(2\pi)^d}
              \log\left[p_0^2-E_{\vec p}^2 -\phi_0^2\right] 
     =-\int\frac{d^dp}{(2\pi)^d}\, \sqrt{E^2_{\vec p}+\phi_0^2}\ .
\ee
The integral can be computed analytically. Expanding to first order 
in $\epsilon=4-d$ we get 
\be
\label{VLOa}
 V_0 =  \frac{\phi_0}{3}\left[ 
   1 + \frac{7-3(\gamma+\log(2))}{6}\, \epsilon \right]
  \left(\frac{m\phi_0}{2\pi}\right)^{d/2} .
\ee
The $\mu$ insertion is given by 
\be
 V_1  = \mu\int\frac{d^dp}{(2\pi)^d}
     \frac{E_{\vec p}}{\sqrt{E^2_{\vec p}+\phi_0^2}} .
\ee
Again, the integral can be computed analytically. The result is 
\be
 V_1 = -\frac{\mu}{\epsilon}\left[ 
   1 + \frac{1-2(\gamma-\log(2))}{4}\, \epsilon \right]
  \left(\frac{m\phi_0}{2\pi}\right)^{d/2} 
\ee
Nishida and Son also computed the two-loop contribution shown 
in Fig.~\ref{fig_veff_eps}. The result is 
\be 
V_2 = -C\epsilon  \left(\frac{m\phi_0}{2\pi}\right)^{d/2} ,
\ee
where $C\simeq 0.14424$. We can now determine the minimum of the 
effective potential. We find
\be 
\phi_0 = \frac{2\mu}{\epsilon}\,\left[ 1 + 
    (3C-1+\log(2))\,\epsilon + O(\epsilon^2) \right] .
\ee
The value of $V=V_0+V_1+V_2$ at $\phi_0$ determines the pressure
and $n=\partial P/\partial \mu$ gives the density. We find
\be 
\label{n_int}
 n = \frac{1}{\epsilon}\,\left[ 1 - \frac{1}{4}
  \left( 2\gamma-1-\log(2) \right) + O(\epsilon^2) \right]  
 \left(\frac{m\phi_0}{2\pi}\right)^{d/2}.
\ee
For comparison, the density of a free Fermi gas in $d$ dimensions is 
\be 
\label{n_free}
 n = \frac{2}{(4\pi)^{d/2}} 
    \frac{k_F^d}{\Gamma\left(1+\frac{d}{2}\right)}.
\ee
This equation determines the relation between $\epsilon_F\equiv 
k_F^2/(2m)$ and the density. We get 
\be 
\label{e_f}
\epsilon_F = \frac{2\pi}{m} \left[\frac{n}{2}
  \Gamma\left(\frac{d}{2}+1\right)\right]^{2/d}.
\ee
We determine $\epsilon_F$ for the interacting gas by inserting 
$n$ from equ.~(\ref{n_int}) into equ.~(\ref{e_f}). The universal   
parameter is $\xi=\mu/\epsilon_F$ . We find 
\be 
\xi = \frac{1}{2}\epsilon^{3/2} 
      + \frac{1}{16}\epsilon^{5/2}\log(\epsilon)
      - 0.025\epsilon^{5/2} + \ldots 
    = 0.475 \hspace{0.25cm} (\epsilon=1),
\ee
which agrees quite well with the result of fixed node quantum 
Monte Carlo calculations. The calculation has been extended to $O(
\epsilon^{7/2})$ by Arnold et al.~\cite{Arnold:2006fr}. Unfortunately,
the next term is very large and it appears necessary to combine the 
expansion in $4-\epsilon$ dimensions with a $2+\epsilon$ expansion 
in order to extract useful results. The $\epsilon$ expansion has also 
been applied to the calculation of the gap \cite{Nishida:2006br}, the 
critical temperature \cite{Nishida:2006rp} and the critical chemical 
potential imbalance \cite{Rupak:2006et,Nishida:2006eu}.

%%%%%%%%%%%%%%%%%%%%%%%%%%%%%%%%%%%%%%%%%%%%%%%%%%%%%%%%%%%%%%%%%%%%
\section{QCD and its symmetries}
\subsection{Introduction}
\label{sec_qcd}
%%%%%%%%%%%%%%%%%%%%%%%%%%%%%%%%%%%%%%%%%%%%%%%%%%%%%%%%%%%%%%%%%%%%

 Before we discuss QCD at large baryon density we would like to 
provide a quick review of QCD and the symmetries of QCD. The elementary 
degrees of freedom are quark fields $\psi^a_{\alpha,f}$ and gluons 
$A_\mu^a$. Here, $a$ is color index that transforms in the fundamental 
representation for fermions and in the adjoint representation for gluons. 
Also, $f$ labels the quark flavors $u,d,s,c,b,t$. In practice, we will 
focus on the three light flavors up, down and strange. The QCD lagrangian 
is 
\be
\label{l_qcd}
 {\cal L } = \sum_f^{N_f} \bar{\psi}_f ( i\Dslash - m_f) \psi_f
  - \frac{1}{4} G_{\mu\nu}^a G_{\mu\nu}^a,
\ee
where the field strength tensor is defined by 
\be
 G_{\mu\nu}^a = \partial_\mu A_\nu^a - \partial_\nu A_\mu^a
  + gf^{abc} A_\mu^b A_\nu^c,
\ee
and the covariant derivative acting on quark fields is 
\be
 i\Dslash \psi = \gamma^\mu \left(
 i\partial_\mu + g A_\mu^a \frac{\lambda^a}{2}\right) \psi.
\ee
QCD has a number of remarkable properties. Most remarkably, even though 
QCD accounts for the rich phenomenology of hadronic and nuclear physics, 
it is an essentially parameter free theory. To first approximation, the 
masses of the light quarks $u,d,s$ are too small to be important, while 
the masses of the heavy quarks $c,b,t$ are too heavy. If we set the masses 
of the light quarks to zero and take the masses of the heavy quarks to be 
infinite then the only parameter in the QCD lagrangian is the coupling 
constant, $g$. Once quantum corrections are taken into account $g$ becomes 
a function of the scale at which it is measured. If the scale is large then 
the coupling is small, but in the infrared the coupling becomes large. This 
is the famous phenomenon of asymptotic freedom. Since the coupling depends 
on the scale the dimensionless parameter $g$ is traded for a dimensionful 
scale parameter $\Lambda_{QCD}$. Since $\Lambda_{QCD}$ is the only 
dimensionful quantity in QCD with massless fermions it is not really a 
parameter of QCD, but reflects our choice of units. In standard units, 
$\Lambda_{QCD} \simeq 200\,{\rm MeV} \simeq 1\,{\rm fm}^{-1}$. 

 Another important feature of the QCD lagrangian are its symmetries. First 
of all, the lagrangian is invariant under local gauge transformations $U(x)
\in SU(3)_c$
\be 
\psi(x) \to U(x)\psi(x),\hspace{1cm}
A_\mu(x) \to U(x)A_\mu U^\dagger (x)
 + iU(x)\partial_\mu U^\dagger(x),
\ee
where $A_\mu= A_\mu^a(\lambda^a/2)$. In the QCD ground state at zero 
temperature and density the local color symmetry is confined. This implies 
that all excitations are singlets under the gauge group. 

 The dynamics of QCD is completely independent of flavor. This implies that 
if the masses of the quarks are equal, $m_u=m_d=m_s$, then the theory is 
invariant under arbitrary flavor rotations of the quark fields 
\be
 \psi_f\to V_{fg}\psi_g,
\ee
where 
$V\in SU(3)$. This is the well known flavor (isospin) symmetry of the 
strong interactions. If the quark masses are not just equal, but equal to 
zero, then the flavor symmetry is enlarged. This can be seen by defining 
left and right-handed fields
\be
  \psi_{L,R} = \frac{1}{2} (1\pm \gamma_5) \psi .
\ee
In terms of $L/R$ fields the fermionic lagrangian is
\be
 {\cal L} =  \bar{\psi}_L (i\Dslash) \psi_L
    +\bar{\psi}_R (i\Dslash) \psi_R + 
   \bar{\psi}_L M \psi_R + \bar{\psi}_R M\psi_L ,
\ee
where $M = {\rm diag}(m_u,m_d,m_s)$. We observe that if quarks are massless, 
$m_u=m_d=m_s=0$, then there is no coupling between left and right handed 
fields. As a consequence, the lagrangian is invariant under independent
flavor transformations of the left and right handed fields.
\be
 \psi_{L,f}\to L_{fg}\psi_{L,g}, \hspace{1cm}
 \psi_{R,f}\to R_{fg}\psi_{R,g},
\ee
where $(L,R)\in SU(3)_L\times SU(3)_R$. In the real world, of course, the 
masses of the up, down and strange quarks are not zero. Nevertheless, since 
$m_u,m_d\ll m_s < \Lambda_{QCD}$ QCD has an approximate chiral symmetry. 

 In the QCD ground state at zero temperature and density the flavor 
symmetry is realized, but the chiral symmetry is spontaneously broken by 
a quark-anti-quark condensate $\langle \bar\psi_L\psi_R +\bar\psi_R\psi_L
\rangle$. As a result, the observed hadrons can be approximately assigned 
to representations of the $SU(3)_V$ flavor group, but not to representations 
of $SU(3)_L\times SU(3)_R$. Nevertheless, chiral symmetry has important 
implications for the dynamics of QCD at low energy. Goldstone's theorem
implies that the breaking of $SU(3)_L\times SU(3)_R\to SU(3)_V$ is associated 
with the appearance of an octet of (approximately) massless pseudoscalar 
Goldstone bosons. Chiral symmetry places important restrictions on the
interaction of the Goldstone bosons. These constraints are obtained most 
easily from the low energy effective chiral lagrangian. At leading order 
we have
\be
\label{l_chpt}
{\cal L} = \frac{f_\pi^2}{4} {\rm Tr}\left[
 \partial_\mu\Sigma\partial^\mu\Sigma^\dagger\right] 
 +\Big[ B {\rm Tr}(M\Sigma^\dagger) + h.c. \Big]
+ \ldots, 
\ee
where $\Sigma=\exp(i\phi^a\lambda^a/f_\pi)$ is the chiral field, $f_\pi$ is 
the pion decay constant and $M$ is the mass matrix. Expanding $\Sigma$ in 
powers of the pion, kaon and eta fields $\phi^a$ we can derive the leading 
order chiral perturbation theory results for Goldstone boson scattering and 
the coupling of Goldstone bosons to external fields. Higher order corrections 
originate from loops and higher order terms in the effective lagrangian.

 Finally, we observe that the QCD lagrangian has two $U(1)$ symmetries,
\bea
U(1)_B: \hspace{1cm}& \psi_L\to e^{i\phi}\psi_L, \hspace{1cm}& 
     \psi_R\to e^{i\phi}\psi_R \\
U(1)_A: \hspace{1cm}& \psi_L\to e^{i\alpha}\psi_L,\hspace{1cm} & 
     \psi_R\to e^{-i\alpha}\psi_R .
\eea
The $U(1)_B$ symmetry is exact even if the quarks are not massless. 
Superficially, it appears that the $U(1)_A$ symmetry is explicitly broken 
by the quark masses and spontaneously broken by the quark condensate. 
However, there is no Goldstone boson associated with spontaneous $U(1)_A$ 
breaking. The reason is that at the quantum level the $U(1)_A$ symmetry is 
broken by an anomaly. The divergence of the $U(1)_A$ current is given by
\be 
\partial^\mu j_\mu^5 = \frac{N_f g^2}{16\pi^2}
 G^a_{\mu\nu}\tilde{G}^a_{\mu\nu},
\ee
where $\tilde{G}^a_{\mu\nu}=\epsilon_{\mu\nu\alpha\beta}
G^a_{\alpha\beta}/2$ is the dual field strength tensor.

%%%%%%%%%%%%%%%%%%%%%%%%%%%%%%%%%%%%%%%%%%%%%%%%%%%%%%%%%%%%%%%%%%%%
\subsection{QCD at finite density}
\label{sec_dqcd}
%%%%%%%%%%%%%%%%%%%%%%%%%%%%%%%%%%%%%%%%%%%%%%%%%%%%%%%%%%%%%%%%%%%%

 In the real world the quark masses are not equal and the only exact 
global symmetries of QCD are the $U(1)_f$ flavor symmetries associated 
with the conservation of the number of up, down, and strange quarks. 
If we take into account the weak interactions then flavor is no longer 
conserved and the only exact symmetries are the $U(1)_B$ of baryon 
number and the $U(1)_Q$ of electric charge. 

 In the following we study hadronic matter at non-zero baryon density. 
We will mostly focus on systems at non-zero baryon chemical potential 
but zero electron $U(1)_Q$ chemical potential. We should note that in 
the context of neutron stars we are interested in situations when the 
electric charge, but not necessarily the electron chemical potential, 
is zero. Also, if the system is in equilibrium with respect to strong, 
but not to weak interactions, then non-zero flavor chemical potentials 
may come into play. 

\begin{figure}[t]
\bc\includegraphics[width=9.0cm]{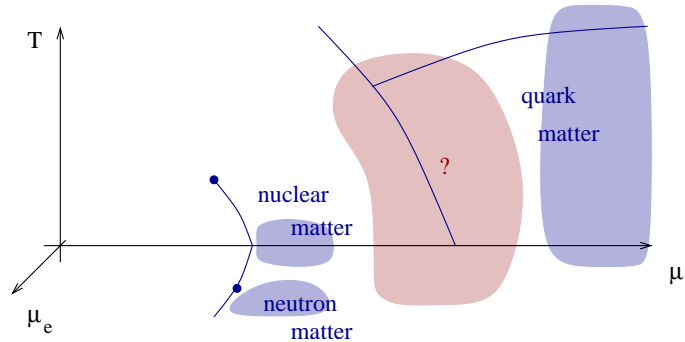}\ec
\caption{\label{fig_phase_1}
Schematic phase diagram of hadronic matter as a function of the
baryon and electron chemical potentials and temperature.}
\end{figure}

The partition function of QCD at non-zero baryon chemical 
potential is given by 
\be 
Z = \sum_i \exp\left(-\frac{E_i-\mu N_i}{T}\right),
\ee
where $i$ labels all quantum states of the system, $E_i$ and $N_i$ are 
the energy and baryon number of the state $i$. If the temperature and 
chemical potential are both zero then only the ground state contributes 
to the partition function. All other states give contributions that are 
exponentially small if the  volume of the system is taken to infinity. 
In QCD there is a mass gap for states that carry baryon number. As a 
consequence there is an onset chemical potential
\be 
\mu_{c}=\min_i (E_i/N_i),
\ee 
such that the partition function is independent of $\mu$ for $\mu<\mu_{c}$. 
For $\mu>\mu_{c}$ the baryon density is non-zero. If the chemical potential 
is just above the onset chemical potential we can describe QCD, to first 
approximation, as a dilute gas of non-interacting nucleons. In this 
approximation $\mu_{c}=m_N$. Of course, the interaction between nucleons 
cannot be neglected. Without it, we would not have stable nuclei. As a 
consequence, nuclear matter is self-bound and the energy per baryon in 
the ground state is given by
\be 
\frac{E_N}{N}-m_N \simeq -15\,{\rm MeV}.
\ee
The onset transition is a first order transition at which the baryon 
density jumps from zero to nuclear matter saturation density, $\rho_0
\simeq 0.14\,{\rm fm}^{-3}$. The first order transition continues into 
the finite temperature plane and ends at a critical endpoint at 
$T=T_c\simeq 10$ MeV, see Fig.~\ref{fig_phase_1}. 

%%%%%%%%%%%%%%%%%%%%%%%%%%%%%%%%%%%%%%%%%%%%%%%%%%%%%%%%%%%%%%%%%%%%
%%%%%%%%%%%%%%%%%%%%%%%%%%%%%%%%%%%%%%%%%%%%%%%%%%%%%%%%%%%%%%%%%%%%

 Nuclear matter is a complicated many-body system and, unlike the 
situation at zero density and finite temperature, there is little 
information from numerical simulations on the lattice. This is related 
to the so-called 'sign problem'. At non-zero chemical potential the 
euclidean fermion determinant is complex and standard Monte-Carlo 
techniques based on importance sampling fail. Recently, some progress 
has been made in simulating QCD for small $\mu$ and $T\simeq T_c$
\cite{Fodor:2001pe,deForcrand:2002ci,Allton:2002zi}, but the regime 
of small temperature remains inaccessible. 

 In neutron stars there is a non-zero electron chemical potential 
and matter is neutron rich. Pure neutron matter has positive pressure
and is stable at arbitrarily low density. As we emphasized in 
Sect.~\ref{sec_uni} dilute neutron matter has universal properties
that can be explored using atomic systems. As the density increases
three and four-body interactions as well as short range forces 
become more important and effective field theory methods are no 
longer applicable. 

 If the density is much larger than nuclear matter saturation density, 
$\rho\gg\rho_0$, we expect the problem to simplify. In this regime it 
is natural to use a system of non-interacting quarks as a starting point 
\cite{Collins:1974ky}. The low energy degrees of freedom are quark 
excitations and holes in the vicinity of the Fermi surface. Since the 
Fermi momentum is large, asymptotic freedom implies that the interaction 
between quasi-particles is weak. We shall see that this does not 
imply that the phase diagram is simple, but it does imply that the 
phase structure can be studied in a systematic fashion.

%%%%%%%%%%%%%%%%%%%%%%%%%%%%%%%%%%%%%%%%%%%%%%%%%%%%%%%%%%%%%%%%%%%%%%%
\section{Effective field theory near the Fermi surface}
\subsection{High density effective theory}
\label{sec_hdet}
%%%%%%%%%%%%%%%%%%%%%%%%%%%%%%%%%%%%%%%%%%%%%%%%%%%%%%%%%%%%%%%%%%%%%%%

 The QCD Lagrangian in the presence of a chemical potential is given by
\be
\label{qcd}
 {\mathcal L} = \bar\psi \left( i\Dslash +\mu\gamma_0 -M \right)\psi
 -\frac{1}{4}G^a_{\mu\nu}G^a_{\mu\nu},
\ee
where $D_\mu=\partial_\mu+igA_\mu$ is the covariant derivative, $M$ 
is the mass matrix and $\mu$ is the baryon chemical potential. If the
baryon chemical potential is large, $\mu\gg\Lambda_{QCD}$, then we 
expect the effective coupling to be small and weak coupling methods 
to be applicable. We shall see, however, that the weak coupling 
expansion is not a simple expansion in the number of loops. Effective
field theory methods are useful in constructing a systematic weak 
coupling expansion. 

 The main observation is that the relevant low energy degrees of freedom 
are particle and hole excitations in the vicinity of the Fermi surface. 
We shall describe these excitations in terms of the field $\psi_v(x)$, 
where $v$ is the Fermi velocity. At tree level, the quark field $\psi$ 
can be decomposed as $\psi=\psi_{v,+}+\psi_{v,-}$ where $\psi_{v,\pm}=
P_{v,\pm}\psi$ with $P_{v,\pm}=\frac{1}{2}(1\pm\vec{\alpha}\cdot\hat{v})
\psi$. Note that $P_{v,\pm}$ is a projector on states with positive/negative 
energy. To leading order in $1/\mu$ we can eliminate the field $\psi_-$ 
using its equation of motion. The lagrangian for the $\psi_+$ field is 
given by \cite{Hong:2000tn,Hong:2000ru,Nardulli:2002ma}
\be
\label{l_hdet}
{\cal L} = \psi_{v}^\dagger \left( iv\cdot D - \frac{D_\perp^2}{2\mu}
 -\frac{g\sigma_{\mu\nu}G_\perp^{\mu\nu}}{4\mu} + \ldots \right) \psi_{v}
 -\frac{1}{4}G^a_{\mu\nu} G^a_{\mu\nu} + \ldots .
\ee
with $v_\mu=(1,\vec{v})$. Note that $v$ labels patches on the Fermi 
surface, and that the number of these patches grows as $\mu^2$. The 
leading order $v\cdot D$ interaction does not connect quarks with 
different $v$, but soft gluons can be exchanged between quarks in 
different patches. In addition to that, there are four, six,  
etc.~fermion operators that contain fermion fields with different 
velocity labels. These operators are constrained by the condition 
that the sum of the velocities has to be zero. 

%%%%%%%%%%%%%%%%%%%%%%%%%%%%%%%%%%%%%%%%%%%%%%%%%%%%%%%%%%%%%%%%%%%%
\begin{figure}[t]
\bc\includegraphics[width=7.5cm]{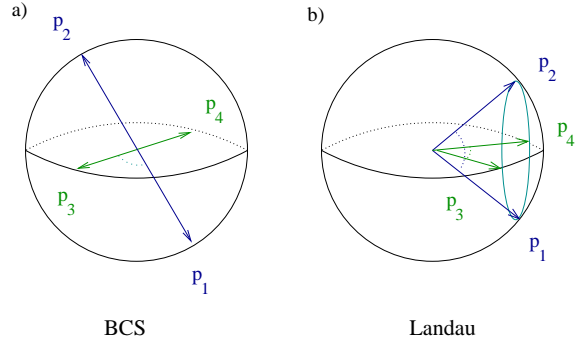}\ec
\caption{\label{fig_fskin}
Kinematics of four-fermion operators in the effective theory.}
\end{figure}
%%%%%%%%%%%%%%%%%%%%%%%%%%%%%%%%%%%%%%%%%%%%%%%%%%%%%%%%%%%%%%%%%%%%

 In the case of four-fermion operators there are two kinds of interactions 
that satisfy this constraint, see Fig.~\ref{fig_fskin}. The first possibility 
is that both the incoming and outgoing fermion momenta are back-to-back. 
This corresponds to the BCS interaction
\be
\label{c_bcs}
{\cal L}=  \frac{1}{\mu^2}\sum_{v',\Gamma,\Gamma'}
  V_l^{\Gamma\Gamma'} R_l^{\Gamma\Gamma'}(\vec{v}\cdot\vec{v}')
    \Big(\psi_{v} \Gamma \psi_{-v}\Big)
   \Big(\psi^\dagger_{v'}\Gamma'\psi^\dagger_{-v'}\Big),
\ee
where $\vec{v}\cdot\vec{v}'=\cos\theta$ is the scattering angle, 
$R_l^{\Gamma\Gamma'}(x)$ is a set of orthogonal 
polynomials, and $\Gamma,\Gamma'$ determine the color, flavor and spin 
structure. The second possibility is that the final momenta are 
equal to the initial momenta up to a rotation around the axis
defined by the sum of the incoming momenta. The relevant
four-fermion operator is
\be
\label{c_flp}
{\cal L}=  \frac{1}{\mu^2}\sum_{v',\Gamma,\Gamma'}
  F_l^{\Gamma\Gamma'} R_l^{\Gamma\Gamma'}(\vec{v}\cdot\vec{v}')
    \Big(\psi_{v} \Gamma \psi_{v'}\Big)
   \Big(\psi^\dagger_{\tilde{v}}\Gamma'\psi^\dagger_{\tilde{v}'}
  \Big).
\ee
In a system with short range interactions only the quantities
$F_l(0)$ are known as Fermi liquid parameters.

%%%%%%%%%%%%%%%%%%%%%%%%%%%%%%%%%%%%%%%%%%%%%%%%%%%%%%%%%%%%%%%%%%%%%%%%%
\subsection{Hard Loops}
\label{sec_hloop}
%%%%%%%%%%%%%%%%%%%%%%%%%%%%%%%%%%%%%%%%%%%%%%%%%%%%%%%%%%%%%%%%%%%%%%%%%

 The effective field theory expansion is complicated by the fact that 
the number of patches $N_v\sim \mu^2/\Lambda^2$ grows with the chemical 
potential. This implies that some higher order contributions that 
are suppressed by $1/\mu^2$ can be enhanced by powers of $N_v$. The 
natural solution to this problem is to sum the leading order diagrams 
in the large $N_v$ limit \cite{Schafer:2004zf}. For gluon $n$-point
functions this corresponds to the well known hard dense loop 
approximation \cite{Braaten:1989mz,Blaizot:1993bb,Manuel:1995td}. 

%%%%%%%%%%%%%%%%%%%%%%%%%%%%%%%%%%%%%%%%%%%%%%%%%%%%%%%%%%%%%%%%%%%%
\begin{figure}[t]
\bc\includegraphics[width=11.5cm]{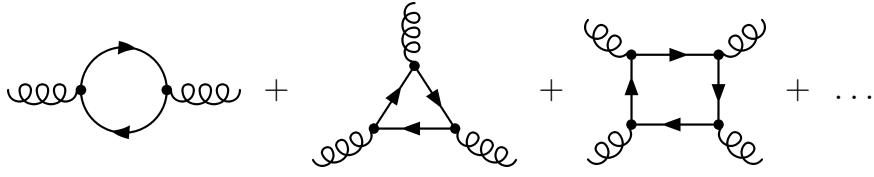}\ec
\caption{\label{fig_hdl}
Hard dense loop contribution to gluon $n$-point functions.}
\end{figure}
%%%%%%%%%%%%%%%%%%%%%%%%%%%%%%%%%%%%%%%%%%%%%%%%%%%%%%%%%%%%%%%%%%%%

 The simplest example is the gluon two point function. At leading 
order in $g$ and $1/\mu$ we have 
\be 
\Pi^{ab}_{\mu\nu}(p) = 2g^2N_f\frac{\delta^{ab}}{2}
  \sum_{\vec{v}} v_\mu v_\nu \int \frac{d^4k}{(2\pi)^4} 
   \frac{1}{(k_0-l_k)(k_0+p_0-l_{k+p})},
\ee
where $l_k=\vec{v}\cdot \vec{k}$. We note that taking the momentum
of the external gluon to zero automatically selects forward scattering.
We also observe that the gluon can interact with fermions of any 
Fermi velocity so that the polarization function involves a sum over 
all patches. After performing the $k_0$ integration we get 
\be 
\Pi^{ab}_{\mu\nu}(p) = 
 2g^2N_f\frac{\delta^{ab}}{2}
  \sum_{\vec{v}} v_\mu v_\nu \int \frac{d^2l_\perp}{(2\pi)^2}
       \int \frac{dl_k}{2\pi} \frac{l_p}{p_0-l_p}
       \frac{\partial n_k}{\partial l_k},
\ee
where $n_k$ is the Fermi distribution function. We note that the $l_k$ 
integration is automatically restricted to small momenta. The integral 
over the transverse momenta $l_\perp$, on the other hand, diverges 
quadratically with the cutoff $\Lambda_\perp$. We observe, however, that 
the sum over patches and the integral over $l_\perp$ can be combined into 
an integral over the entire Fermi surface
\be
\label{hard_int}
 \frac{1}{2\pi}\sum_{\vec{v}}\int\frac{d^2l_\perp}{(2\pi)^2}
 =\frac{\mu^2}{2\pi^2}\int \frac{d\Omega}{4\pi}.
\ee
This means that the transverse momentum integral is extended all the 
way up to $\mu$. Because the energy of the fermions is small but the 
loop momentum is large the integral is referred to as a hard dense
loop. We find
\be
\label{pi_hdet2}
\Pi^{ab}_{\mu\nu}(p) = 2m^2\delta^{ab} \int\frac{d\Omega}{4\pi}
  v_\mu v_\nu \left\{ 1-\frac{p_0}{p_0-l_p}
 \right\},
\ee
where we have defined the effective gluon mass $m^2=N_F g^2\mu^2/(4\pi^2)$.
This result has the same structure as the non-relativistic expression 
given in equ.~(\ref{pi_ij}), but the tadpole contribution is missing. 
As a consequence, equ.~(\ref{pi_hdet2}) is not transverse. In the 
relativistic theory the tadpole contribution originates from the
$D_\perp^2/(2\mu)$ in the effective lagrangian. The tadpole is proportional 
to the total density and corresponds to a counterterm  \cite{Hong:2000tn}
\be
{\cal L}= \frac{1}{2} m^2 \int\frac{d\Omega}{4\pi}
 (\vec{A}_\perp)^2.
\ee
Putting everything together we find
\be
\label{pi_hdet3}
\Pi^{ab}_{\mu\nu}(p) = 2m^2\delta^{ab} \int\frac{d\Omega}{4\pi}
   \left\{ \delta_{\mu 0}\delta_{\nu 0} -
   \frac{v_\mu v_\nu p_0}{p_0-l_p} \right\}.
\ee
The gluonic three-point function shown in Fig.~\ref{fig_hdl}b can be 
computed in the same fashion. We find
\be 
\label{gam_hdet}
 \Gamma^{abc}_{\mu\nu\alpha}(p,q,r) = igf^{abc} 2m^2 
   \int \frac{d\Omega}{4\pi} v_\mu v_\alpha v_\beta \left\{ 
    \frac{q_0}{(q\cdot v)(p\cdot v)}-\frac{r_0}{(r\cdot v)(p\cdot v)}
 \right\},
\ee
where $p,q,r$ are the incoming gluon momenta ($p+q+r=0$). We note that 
in the case of the three point function, as well as in all higher $n$-point 
functions, there are no tadpole or counterterm contributions. There is a 
simple generating functional for these loop integrals which is known as 
the hard dense loop (HDL) effective action  \cite{Braaten:1991gm}
\be
\label{S_hdl}
{\cal L}_{HDL} = -\frac{m^2}{2}\sum_v \,G^a_{\mu\alpha}
  \frac{v^\alpha v^\beta}{(v\cdot D)^2} G^a_{\mu\beta}.
\ee
This is a gauge invariant, but non-local, effective lagrangian.

%%%%%%%%%%%%%%%%%%%%%%%%%%%%%%%%%%%%%%%%%%%%%%%%%%%%%%%%%%%%%%%%%%%%%%%
\subsection{Non-Fermi liquid effective field theory}
\label{sec_nfl}
%%%%%%%%%%%%%%%%%%%%%%%%%%%%%%%%%%%%%%%%%%%%%%%%%%%%%%%%%%%%%%%%%%%%%%%

 In this Section we shall study the effective field theory in the 
regime $\omega<m$ where $\omega$ is the excitation energy and $m$ 
is the effective gluon mass \cite{Schafer:2005mc}. In the previous 
section we argued that hard dense loops have to be resummed in order 
to obtain a consistent low energy expansion. The effective lagrangian 
is given by
\be
\label{l_udet}
 {\cal L} = \psi_{v}^\dagger  \left(  i v\cdot D -
                   \frac{D_\perp^2}{2\mu}   \right) \psi_{v}
    - \frac{1}{4}G^a_{\mu\nu}G^a_{\mu\nu} 
    + {\cal L}_{HDL} + {\cal L}_{4f} + \ldots \, .
\ee 
Since electric fields are screened the interaction at low energies 
is dominated by the exchange of magnetic gluons. The transverse 
gauge boson propagator is
\be
D_{ij}(k) = -\frac{i(\delta_{ij}-\hat{k}_i\hat{k}_j)}
      {k_0^2-\vec{k}^2+i\frac{\pi}{2}m^2 \frac{k_0}{|\vec{k}|}} ,
\ee
where we have assumed that $|k_0|<|\vec{k}|$. We observe that the
propagator becomes large in the regime $|k_0|\sim |\vec{k}|^3/m^2$. 
If the energy is small, $|k_0|\ll m$, then the typical energy is 
much smaller than the typical momentum,
\be
\label{ld_kin}
 |\vec{k}| \sim (m^2 |k_0|)^{1/3} \gg |k_0| .
\ee
This implies that the gluon is very far off its energy shell and not 
a propagating state. We can compute loop diagrams containing quarks 
and transverse gluons by picking up the pole in the quark propagator, 
and then integrate over the cut in the gluon propagator using the 
kinematics dictated by equ.~(\ref{ld_kin}). In order for a quark to 
absorb the large momentum carried by a gluon and stay close to the 
Fermi surface the gluon momentum has to be transverse to the momentum 
of the quark. This means that the term $k_\perp^2/(2\mu)$ in the quark 
propagator is relevant and has to be kept at leading order. Equation 
(\ref{ld_kin}) shows that $k_\perp^2/(2\mu)\gg k_0$ as $k_0\to 0$. This 
means that the pole of the quark propagator is governed by the condition 
$k_{||}\sim k_\perp^2/(2\mu)$. We find
\be
\label{ld_reg}
 k_\perp \sim g^{2/3}\mu^{2/3}k_0^{1/3},\hspace{0.5cm}
 k_{||}  \sim g^{4/3}\mu^{1/3}k_0^{2/3}.
\ee
In the low energy regime propagators and vertices can be simplified
even further. The quark and gluon propagators are
\be
   S_{\alpha\beta}(p) = \frac{i\delta_{\alpha\beta}}
       {p_0-p_{||}-\frac{p_\perp^2}{2\mu}
              +i\epsilon {\it sgn}(p_0)},
\ee\be
   D_{ij}(k) = \frac{i\delta_{ij}}
       {k_\perp^2-i\frac{\pi}{2}m^2\frac{k_0}{k_\perp}},
\ee
and the quark gluon vertex is $gv_i(\lambda^a/2)$. Higher order 
corrections can be found by expanding the quark and gluon propagators 
as well as the HDL vertices in powers of the small parameter $\epsilon
\equiv (k_0/m)$.

%%%%%%%%%%%%%%%%%%%%%%%%%%%%%%%%%%%%%%%%%%%%%%%%%%%%%%%%%%%
\begin{figure}[t]
\includegraphics[width=4.25cm]{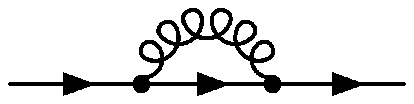}\hspace{-0.75cm}
\includegraphics[width=4.25cm]{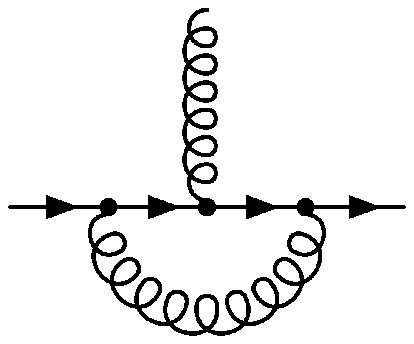}\hspace{-0.75cm}
\includegraphics[width=4.25cm]{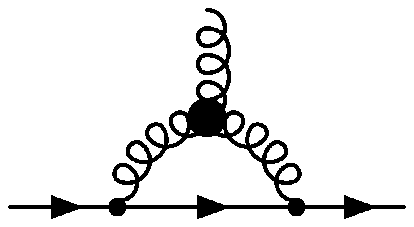}
\caption{One-loop contributions to the quark self energy
and the quark-gluon vertex. The black blob in the third diagram
denotes the HDL gluon three point function. In the magnetic regime
the graphs scale as $\omega\log(\omega)$, $\omega^{1/3}$
and $\omega^{2/3}$, respectively. }
\label{fig_mag}
\end{figure}
%%%%%%%%%%%%%%%%%%%%%%%%%%%%%%%%%%%%%%%%%%%%%%%%%%%%%%%%%%%

 The regime in which all momenta, including external ones, satisfy the 
scaling relation (\ref{ld_reg}) is completely perturbative, i.e.~graphs 
with extra loops are always suppressed by extra powers of $\epsilon^{1/3}$. 
One way to see this is to rescale the fields in the effective lagrangian 
so that the kinetic terms are scale invariant under the transformation 
$(x_0,x_{||},x_\perp)$ $\to$ $(\epsilon^{-1}x_0,\epsilon^{-2/3} x_{||},
\epsilon^{-1/3}x_{\perp})$. The scaling behavior of the fields is
$\psi \to \epsilon^{5/6}\psi$ and $A_i \to \epsilon^{5/6} A_i$. We find 
that the scaling dimension of all interaction terms is positive. The quark 
gluon vertex scales as $\epsilon^{1/6}$, the HDL three gluon vertex scales 
as $\epsilon^{1/2}$, and the four gluon vertex scales as $\epsilon$. Since 
higher order diagrams involve at least one pair of quark gluon vertices 
the expansion involves positive powers of $\epsilon^{1/3}$. 

 As a simple example we consider the fermion self energy in the limit 
$p_0\to 0$. The one-loop diagram is 
\bea
\Sigma(p) &=& -ig^2 C_F\int \frac{dk_0}{2\pi}
\int \frac{dk_\perp^2}{(2\pi)^2} \ \frac{k_\perp}{k_\perp^3+i\eta k_0}
 \nonumber \\
 & & \hspace{0.5cm}\mbox{}
  \times\int \frac{dk_{||}}{2\pi} \  \frac{\Theta(p_0+k_0)}{k_{||}+p_{||}
          -(k_\perp+p_\perp)^2/(2\mu)+i\epsilon}, 
\eea
with $C_F=(N_c^2-1)/(2N_c)$ and $\eta=(\pi/2)m^2$. This expression
shows a number of interesting features. First we observe that the 
longitudinal and transverse momentum integrations factorize. The longitudinal 
momentum integral can be performed by picking up the pole in the quark 
propagator. The result is independent of the external momenta and only 
depends on the external energy. The transverse momentum integral is 
logarithmically divergent. We find \cite{Vanderheyden:1996bw,Manuel:2000mk,Brown:2000eh,Boyanovsky:2000bc,Ipp:2003cj}
\be 
\label{sig_m}
\Sigma(p) = \frac{g^2}{9\pi^2}p_0\log\left(\frac{\Lambda}{|p_0|}
\right),
\ee
where $\Lambda$ is a cutoff for the $k_\perp$ integral. We showed
that the logarithmic divergence can be absorbed in the parameters
of the effective theory \cite{Schafer:2004zf}. In order to fix the 
corresponding counterterm we have include electric gluon exchanges
For $k_0\ll m$ the electric gluon propagator is given by 
\be 
\label{d_00}
 D_{00}(k) = -\frac{i}{\vec{k}^2+2m^2}.
\ee
Higher order corrections can be obtained by expanding the full 
HDL expression in powers of $k_0/m$. The electric contribution 
is dominated by large momenta and does not contribute to fractional
powers or logarithms of $k_0$. We get 
\bea
\label{sig_e}
\Sigma(p) &=& ig^2 C_F\int \frac{dk_0}{2\pi}
\int \frac{dk_\perp^2}{(2\pi)^2} \ \frac{1}{k_\perp^2+2m^2}
 \nonumber \\
 & & \hspace{0.5cm} \mbox{}\times
 \int \frac{dk_{||}}{2\pi} \  \frac{\Theta(p_0+k_0)}{k_{||}+p_{||}
          -(k_\perp+p_\perp)^2/(2\mu)+i\epsilon}.
\eea
This contribution scales as $p_0\log(\Lambda/m)$. The logarithm 
of the cutoff cancels the logarithm in equ.~(\ref{sig_m}). We
get \cite{Ipp:2003cj}
\be
\label{sig}
\Sigma(p_0) = \frac{g^2}{9\pi^2} \left[ 
  p_0 \log\left( \frac{4\sqrt{2}m}{\pi|p_0|} \right)
  + p_0 + i\frac{\pi}{2}|p_0| \right].
\ee
Finally, there are contributions from the hard regime in which both 
the energy and the momentum of the gluon are large, $k_0\sim |\vec{k}|
\geq m$. This regime corresponds to the HDL term in the fermion self 
energy \cite{Manuel:1995td,Schafer:2003jn}. The HDL term gives an
$O(g^2)$ correction to the low energy parameters $v_F$ and $\delta\mu$.

 The logarithmic term in the fermion self energy leads to a breakdown 
of Fermi liquid theory. The quasi-particle  velocity
\be 
 v(p_0) = \frac{1}{1+\Sigma'(p_0)}
\ee
and the wave function renormalization go to zero logarithmically 
as the quasi-particle energy goes to zero. One physical consequence 
of this behavior is an anomalous $T\log(T)$ term in the specific
heat \cite{Holstein:1973,Ipp:2003cj}. The effective theory can also 
be used to study perturbative corrections in other quantities. We find, 
in particular, a QCD version of Migdal's theorem. Migdal showed that 
vertex corrections to the electron-phonon coupling are suppressed
by the ratio of the electron mass to the mass of the positive ions
\cite{Abrikosov:1963}. In the Landau damping regime of QCD loop 
corrections to the quark-gluon vertex are suppressed by powers 
of $\epsilon^{1/3}$. 

%%%%%%%%%%%%%%%%%%%%%%%%%%%%%%%%%%%%%%%%%%%%%%%%%%%%%%%%%%%%%%%%%%%%%%%
\subsection{Color superconductivity}
\label{sec_csc}
%%%%%%%%%%%%%%%%%%%%%%%%%%%%%%%%%%%%%%%%%%%%%%%%%%%%%%%%%%%%%%%%%%%%%%%

 In Sect.~(\ref{sec_bcs}) we showed that the particle-particle 
scattering amplitude in the BCS channel $q(\vec{p}\,)+q(-\vec{p}\,)\to 
q({\vec{p}}^{\,\prime})+q(-{\vec{p}}^{\,\prime})$ is special. The total 
momentum of the pair vanishes and as a consequence loop corrections to 
the scattering amplitude are enhanced. This implies that all ladder 
diagrams have to be summed. Crossed ladders, vertex corrections, etc.~involve 
momenta in the regime $k_\perp\gg k_{||}\gg k_0$ and are perturbative. 

%%%%%%%%%%%%%%%%%%%%%%%%%%%%%%%%%%%%%%%%%%%%%%%%%%%%%%%%%%%%%%%%%%%%
\begin{figure}[t]
\bc\includegraphics[width=11.0cm]{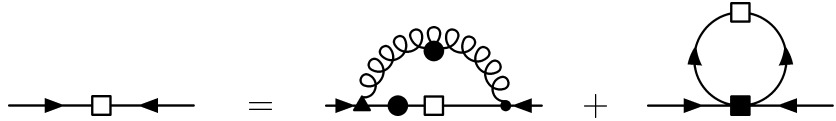}\ec
\caption{\label{fig_bcs_oge}
Gap equation for the superfluid gap in a theory with long 
range interactions.}
\end{figure}
%%%%%%%%%%%%%%%%%%%%%%%%%%%%%%%%%%%%%%%%%%%%%%%%%%%%%%%%%%%%%%%%%%%%

 If the interaction in the particle-particle channel is attractive then 
the BCS singularity leads to the formation of a pair condensate and to 
a gap in the fermion spectrum. The gap can be computed by solving a 
Dyson-Schwinger equation for the anomalous (particle-particle) self 
energy. In QCD the interaction is attractive in the color anti-triplet 
channel. The structure of the gap is simplest in the case of two flavors. 
In that case, there is a unique color anti-symmetric spin zero gap term 
of the form 
\be 
\label{2SC}
 \langle \psi^a_i C\gamma_5 \psi^b_j\rangle  =
        \phi\epsilon^{3ab}\epsilon_{ij}.
\ee
Here, $a,b$ labels color and $i,j$ flavor. The gap equation is given by
\bea
\Delta(p_0) &=& -ig^2 C_A \int \frac{dk_0}{2\pi} 
     \int \frac{dk_\perp^2}{(2\pi)^2} \ 
          \frac{k_\perp}{k_\perp^3+\eta (k_0-p_0)}  \nonumber \\
 & & \hspace{0.5cm}\mbox{}\times
  \int \frac{dk_{||}}{2\pi} \  \frac{\Delta(k_0)}
        {k_0^2+k_{||}^2+\Delta(k_0)^2}, 
\eea
where $C_A=2/3$ is a color factor. Like the normal self energy, the anomalous 
self energy $\Delta(p)$ is dominantly a function of energy. We carry out
the integrals over $k_\perp$ and $k_{||}$ and analytically continue 
to imaginary energy $p_4=ip_0$. The euclidean gap equation is  
\cite{Son:1999uk,Schafer:1999jg,Pisarski:2000tv,Hong:2000fh}
\be
\label{eliash_mel}
\Delta(p_4) = \frac{g^2}{18\pi^2} \int dk_4
 \log\left(\frac{\Lambda_{BCS}}{|p_4-k_4|}\right)
    \frac{\Delta(k_4)}{\sqrt{k_4^2+\Delta(k_4)^2}}.
\ee
The scale $\Lambda_{BCS}$ is sensitive to electric gluon exchange. In 
the anomalous self energy the logarithmic divergence does not cancel 
between magnetic and electric gluon exchanges. The reason is that the 
magnetic contribution is proportional to $\delta_{ij}v_i v_j$ in the 
normal self energy and $\delta_{ij}v_i (-v_j)$ in the anomalous case. 
The logarithmic dependence on the cutoff is absorbed by the BCS four-fermion 
operator. A simple matching calculation gives $\Lambda_{BCS}=256\pi^4
(2/N_f)^{5/2}g^{-5}\mu$ \cite{Schafer:2003jn}. The solution to the gap 
equation was found by Son \cite{Son:1999uk}
\be 
\Delta(x) = \Delta_0\sin\left(\frac{g}{3\sqrt{2}\pi}\, x\right)
\ee
where $x=\log(2\Lambda_{BCS}/(p_4+\omega_p)$ and $\omega_p^2=p_4^2+\Delta_0^2$.
The gap on the Fermi surface is 
\be
\label{gap_oge}
\Delta_0 \simeq 2\Lambda_{BCS}
   \exp\left(-\frac{\pi^2+4}{8}\right)
   \exp\left(-\frac{3\pi^2}{\sqrt{2}g}\right).
\ee
This result is correct up to $O(g)$ corrections to the pre-exponent.
In order to achieve this accuracy the $g^2p_0\log(p_0)$ term 
in the normal self energy, equ.~(\ref{sig_m}), has to be included
in the gap equation \cite{Brown:1999aq,Wang:2001aq,Schafer:2003jn}. 

 The order parameter is slightly more complicated in QCD with $N_f=3$ 
massless flavors. The energetically preferred phase is the 
color-flavor-locked (CFL) phase described by \cite{Alford:1999mk}
\be 
\label{CFL}
 \langle \psi^a_i C\gamma_5 \psi^b_j\rangle  =
   \phi \left(\delta^a_i\delta^b_j-\delta^a_j\delta^b_i\right).
\ee
In the CFL phase there are eight fermions with gap $\Delta_{CFL}$ and 
one fermion with gap $2\Delta_{CFL}$. The CFL gap is given by $\Delta_{CFL}
=2^{-1/3}\Delta_0$ \cite{Schafer:1999fe}. The CFL phase has a number of 
remarkable properties \cite{Alford:1999mk,Schafer:1998ef}. Most notably, 
chiral symmetry is broken in the CFL phase and the low energy spectrum 
contains a flavor octet of pseudoscalar bosons. We shall study the dynamics 
of these Goldstone modes in Sect.~\ref{sec_CFLchi}.

%%%%%%%%%%%%%%%%%%%%%%%%%%%%%%%%%%%%%%%%%%%%%%%%%%%%%%%%%%%%%%%%%%%%%%%
\subsection{Mass terms}
\label{sec_mhdet}
%%%%%%%%%%%%%%%%%%%%%%%%%%%%%%%%%%%%%%%%%%%%%%%%%%%%%%%%%%%%%%%%%%%%%%%

 Mass terms modify the parameters in the effective lagrangian.
These parameters include the Fermi velocity, the effective 
chemical potential, the screening mass, the BCS terms and the
Landau parameters. At tree level the correction to the Fermi
velocity and the chemical potential are given by
\be 
\label{vf_mu}
 v_F= 1-\frac{m^2}{2p_F^2},\hspace{1cm}
\delta \mu =-\frac{m^2}{2p_F}.
\ee
The shift in the Fermi velocity also affects the coupling 
$gv_F$ of a magnetic gluon to quarks. It is important to note 
that at leading order in $g$ this the only mass correction
to the coupling. This is not entirely obvious, as one can 
imagine a process in which the quark emits a gluon, makes 
a transition to a virtual high energy state, and then couples
back to a low energy state by a mass insertion. This process
would give an $O(m/\mu)$ correction to $g$, but it vanishes
in the forward direction \cite{Schafer:2001za}. 

%%%%%%%%%%%%%%%%%%%%%%%%%%%%%%%%%%%%%%%%%%%%%%%%%%%%%%%%%%%%%%%%%%%%
\begin{figure}[t]
\bc\includegraphics[width=8.25cm]{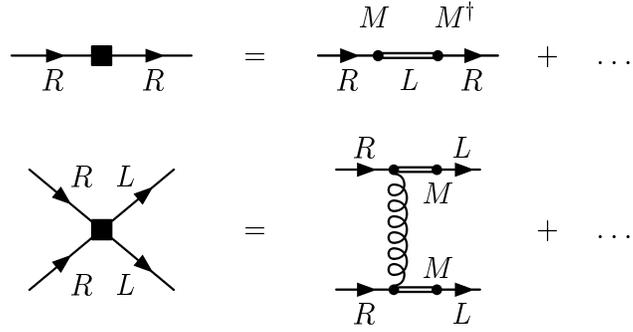}\ec
\caption{\label{fig_hdet_m}
Mass terms in the high density effective theory. The first 
diagram shows a $O(MM^\dagger)$ term that arises from integrating 
out the $\psi_-$ field in the QCD lagrangian. The second
diagram shows a $O(M^2)$ four-fermion operator which arises from 
integrating out $\psi_-$ and hard gluon exchanges.}
\end{figure}
%%%%%%%%%%%%%%%%%%%%%%%%%%%%%%%%%%%%%%%%%%%%%%%%%%%%%%%%%%%%%%%%%%%%

  Quark masses modify quark-quark scattering amplitudes and 
the corresponding Landau and BCS type four-fermion operators. 
Consider quark-quark scattering in the forward direction, 
$v+v'\to v+v'$. At tree level in QCD this process receives
contribution from the direct and exchange graph. In the 
effective theory the direct term is reproduced by the collinear 
interaction while the exchange terms has to be matched against 
a four-fermion operator. The spin-color-flavor symmetric 
part of the exchange amplitude is given by 
\be
\label{m_fl}
{\cal M}(v,v';v,v') =  \frac{C_F}{4N_c N_f}\frac{g^2}{p_F^2}
  \left\{ 1-\frac{m^2}{p_F^2}\frac{x}{1-x} \right\}
\ee
where $C_F=(N_c^2-1)/(2N_c)$ and $x=\hat{v}\cdot\hat{v}'$ is 
the scattering angle. We observe that the amplitude is independent 
of $x$ in the limit $m\to 0$. Mass corrections are singular as 
$x\to 1$. The means that the Landau coefficients $F_l$ contain
logarithms of the cutoff. We note that there is one linear combination 
of Landau coefficients, $F_0-F_1/3$, which is cutoff independent. 

 Equations (\ref{vf_mu}-\ref{m_fl}) are valid for $N_f\geq 1$ degenerate 
flavors. Spin and color anti-symmetric BCS amplitudes require at least 
two different flavors. Consider BCS scattering $v+(-v)\to v'+(-v')$
in the helicity flip channel $L+L\to R+R$. The color-anti-triplet 
amplitude is given by
\be 
\label{m_bcs}
{\cal M}(v,-v;v',-v') = \frac{C_A}{4}\frac{g^2}{p_F^2}
  \frac{m_1m_2}{p_F^2}.
\ee 
where $m_1$ and $m_2$ are the masses of the two quarks and $C_A=(N_c+1)/
(2N_c)$. We observe that the scattering amplitude is independent of the 
scattering angle. This means that at leading order in $g$ and $m$ only 
the s-wave potential $V_0$ is non-zero. 

 In order to match Green functions in the high density effective 
theory to an effective chiral theory of the CFL phase we need to 
generalize our results to a complex mass matrix of the form ${\cal L}=  
-\bar\psi_L M\psi_R - \bar\psi_R M^\dagger \psi_L $, see
Fig.~\ref{fig_hdet_m}. The $\delta\mu$ term is 
\be 
\label{m_kin}
{\mathcal L} = -\frac{1}{2p_F} \left( \psi_{L+}^\dagger MM^\dagger \psi_{L+}
 + \psi_{R+}^\dagger M^\dagger M\psi_{R+} \right).
\ee
and the four-fermion operator in the BCS channel is
\bea
\label{hdet_m}
 {\mathcal L} &=& \frac{g^2}{32p_F^4}
     \left(\psi^{a\,\dagger}_{i,L} C\psi^{b\,\dagger}_{j,L}\right)
     \left(\psi^c_{k,R} C \psi^d_{l,R}\right) 
    \left[(\vec\lambda)^{ac}(\vec\lambda)^{bd}(M)_{ik}(M)_{jl} \right]
    \nonumber \\
 & & \hspace{1cm}\mbox{} 
  \; + \; \left(L\leftrightarrow R, M\leftrightarrow M^\dagger\right)\, .
\eea

%%%%%%%%%%%%%%%%%%%%%%%%%%%%%%%%%%%%%%%%%%%%%%%%%%%%%%%%%%%%%%%%%
\section{Chiral theory of the CFL phase}
\label{sec_CFLchi}
%%%%%%%%%%%%%%%%%%%%%%%%%%%%%%%%%%%%%%%%%%%%%%%%%%%%%%%%%%%%%%%%%

%%%%%%%%%%%%%%%%%%%%%%%%%%%%%%%%%%%%%%%%%%%%%%%%%%%%%%%%%%%%%%%%%
\subsection{Introduction}
\label{sec_CFLchi_intro}
%%%%%%%%%%%%%%%%%%%%%%%%%%%%%%%%%%%%%%%%%%%%%%%%%%%%%%%%%%%%%%%%%

 For energies smaller than the gap the only relevant degrees
of freedom are the Goldstone modes associated with spontaneously
broken global symmetries. The quantum numbers of the Goldstone 
modes depend on the symmetries of the order parameter. In the 
following we shall concentrate on the CFL phase. Goldstone modes
determine the specific heat, transport properties, and the 
response to external fields for temperatures less than $T_c$. 
As we shall see, Goldstone modes also determine the phase 
structure as a function of the quark masses. 

%%%%%%%%%%%%%%%%%%%%%%%%%%%%%%%%%%%%%%%%%%%%%%%%%%%%%%%%%%%%%%%%%
\subsection{Chiral effective field theory}
\label{sec_CFLeft}
%%%%%%%%%%%%%%%%%%%%%%%%%%%%%%%%%%%%%%%%%%%%%%%%%%%%%%%%%%%%%%%%%

 In the CFL phase the pattern of chiral symmetry breaking is identical 
to the one at $T=\mu=0$. This implies that the effective lagrangian has 
the same structure as chiral perturbation theory. The main difference is 
that Lorentz-invariance is broken and only rotational invariance is a 
good symmetry. The effective lagrangian for the Goldstone modes is given 
by \cite{Casalbuoni:1999wu}
\bea
\label{l_cheft}
{\mathcal L}_{eff} &=& \frac{f_\pi^2}{4} {\rm Tr}\left[
 \nabla_0\Sigma\nabla_0\Sigma^\dagger - v_\pi^2
 \partial_i\Sigma\partial_i\Sigma^\dagger \right] 
 +\left[ B {\rm Tr}(M\Sigma^\dagger) + h.c. \right] 
    \nonumber \\ 
 & & \hspace*{0cm}\mbox{} 
     +\left[ A_1{\rm Tr}(M\Sigma^\dagger)
                        {\rm Tr} (M\Sigma^\dagger) 
     + A_2{\rm Tr}(M\Sigma^\dagger M\Sigma^\dagger) \right.
 \nonumber \\[0.1cm] 
  & &   \hspace*{0.5cm}\mbox{}\left. 
     + A_3{\rm Tr}(M\Sigma^\dagger){\rm Tr} (M^\dagger\Sigma)
         + h.c. \right]+\ldots . 
\eea
Here $\Sigma=\exp(i\phi^a\lambda^a/f_\pi)$ is the chiral field, $f_\pi$ 
is the pion decay constant and $M$ is a complex mass matrix. The chiral 
field and the mass matrix transform as $\Sigma\to L\Sigma R^\dagger$ and  
$M\to LMR^\dagger$ under chiral transformations $(L,R)\in SU(3)_L\times 
SU(3)_R$. We have suppressed the singlet fields associated with the breaking 
of the exact $U(1)_V$ and approximate $U(1)_A$ symmetries. 

 At low density the coefficients $f_\pi$, $B$, $A_i,\ldots$ are 
non-perturbative quantities that have to extracted from experiment or 
measured on the lattice. At large density, on the other hand, the chiral 
coefficients can be calculated in perturbative QCD. The pion decay 
constant and the pion velocity can be determined by gauging the 
$SU(3)_L\times SU(3)_R$ symmetry. The covariant derivative $D_\mu
\Sigma=\partial_\mu\Sigma +iW_\mu^L\Sigma-i\Sigma W_\mu^R$ generates
mass terms for the gauge field $W_\mu^{L,R}$, 
\be 
{\cal L} = \frac{f_\pi^2}{4}\left( 
   \frac{1}{2} \left( W_0^L-W_0^R \right)^2
 - \frac{v_\pi^2}{2}\left( W_i^L-W_i^R \right)^2\right).
\ee
The electric and magnetic screening masses in the CFL phase can be
determined as in Sect.~\ref{sec_mei}. The main difference is that in 
the CFL phase there are nine different fermion modes, and that not 
all of these modes have the same gap. There is also mixing between
flavor and color gauge fields. It is easiest to compute the screening
for the color gauge fields. The electric screening mass is
\bea
 \Pi_{00} &=& 
 -2i \int\frac{d^4p}{(2\pi)^4} \left\{ 
 \frac{7}{6}\frac{p_0^2+\epsilon_p^2}
        {(p_0^2-\epsilon_p^2-\Delta_8^2)(p_0^2-\epsilon_p^2-\Delta_8^2)}
 \right.\nonumber\\
 & & \mbox{}
 +\frac{1}{3}\frac{p_0^2+\epsilon_p^2}
        {(p_0^2-\epsilon_p^2-\Delta_8^2)(p_0^2-\epsilon_p^2-\Delta_1^2)}
 - \frac{1}{3}\frac{\Delta_8^2}
        {(p_0^2-\epsilon_p^2-\Delta_8^2)(p_0^2-\epsilon_p^2-\Delta_8^2)}
 \nonumber \\
\label{nf3_chiv}
 & & \left.\hspace{1.75cm}\mbox{}
 - \frac{1}{3}\frac{\Delta_8\Delta_1}
        {(p_0^2-\epsilon_p^2-\Delta_8^2)(p_0^2-\epsilon_p^2-4\Delta_1^2)}
  \right\}.
\eea
The first terms comes from particle-hole diagrams with two octet 
quasi-particles while the second term comes from diagrams with one octet 
and one singlet quasi-particle. There is no coupling of an octet field to 
two singlet particles. The third and fourth term are the corresponding 
contributions from particle-particle and hole-hole pairs. In the CFL
phase $\Delta_1=2\Delta_8\equiv 2\Delta$. The four integrals in 
(\ref{nf3_chiv}) give
\be
 \Pi_{00} = 2\left\{ \frac{7}{6} +\frac{1}{3}
  - \frac{1}{3} - \frac{4\log(2)}{9}\right\}
   \left(\frac{\mu^2}{4\pi^2}\right)
 = \frac{21-8\log(2)}{18} \left( \frac{\mu^2}{2\pi^2} \right)
\ee
The magnetic mass can be computed in the same fashion. As in 
Sect.~\ref{sec_mei} we have to add the contribution from the tadpole
and the structure of the gauge field mass term is $m_D^2(A_0^2-\vec{A}^2
/3)$. Mixing between flavor and color gauge fields was studied in 
\cite{Casalbuoni:1999wu,Bedaque:2001je}. The result is that there 
is no screening for the vector field $W_L+W_R$ and that the screening 
mass for the axial field $W_L-W_R$ is equal to the mass of the 
color gauge field. We conclude that \cite{Son:1999cm} 
\be
\label{cfl_fpi}
f_\pi^2 = \frac{21-8\log(2)}{18} \left(\frac{p_F^2}{2\pi^2} \right), 
\hspace{0.5cm} v_\pi^2=\frac{1}{3}.
\ee
Mass terms are determined by the operators studied in 
Sect.~\ref{sec_mhdet}. We observe that both equ.~(\ref{m_kin})
and (\ref{hdet_m}) are quadratic in $M$. This implies that $B=0$
in perturbative QCD. $B$ receives non-perturbative contributions
from instantons, but these effects are small if the density is
large, see \cite{Schafer:2002ty}.

%%%%%%%%%%%%%%%%%%%%%%%%%%%%%%%%%%%%%%%%%%%%%%%%%%%%%%%%%%%%%%%%%%%%
\begin{figure}[t]
\bc\includegraphics[width=10cm]{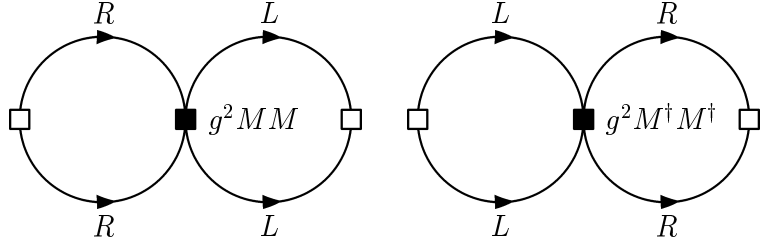}\ec
\caption{\label{fig_4fvac}
Contribution of the $O(M^2)$ BCS four-fermion operator to the 
condensation energy in the CFL phase .}
\end{figure}
%%%%%%%%%%%%%%%%%%%%%%%%%%%%%%%%%%%%%%%%%%%%%%%%%%%%%%%%%%%%%%%%%%%%

  We also note that $X_L=MM^\dagger/(2p_F)$ and $X_R=M^\dagger M/
(2p_F)$ in equ.~(\ref{m_kin}) act as effective chemical potentials 
for left and right-handed fermions, respectively. Formally, the 
effective lagrangian has an $SU(3)_L\times SU(3)_R$ gauge 
symmetry under which $X_{L,R}$ transform as the temporal components
of non-abelian gauge fields. We can implement this approximate gauge 
symmetry in the CFL chiral theory by promoting time derivatives
to covariant derivatives \cite{Bedaque:2001je}, 
\be
\label{mueff}
 \nabla_0\Sigma = \partial_0 \Sigma 
 + i \left(\frac{M M^\dagger}{2p_F}\right)\Sigma
 - i \Sigma\left(\frac{ M^\dagger M}{2p_F}\right) .
\ee
The BCS four-fermion operator in equ.~(\ref{hdet_m}) contributes to 
to the condensation energy in the CFL phase, see Fig.~\ref{fig_4fvac}.
The diagram is proportional to the square of the superfluid 
density 
\bea
\langle \psi^a_{i,L} C\psi^b_{j,L}\rangle
 &=& \left(\delta^a_i\delta^b_j-\delta^a_j\delta^b_i\right)
    \int\frac{d^4p}{(2\pi)^4}
    \frac{\Delta(p_0)}{p^2-\epsilon_p^2-\Delta^2(p_0)}
 \nonumber \\
 &=&  \left(\delta^a_i\delta^b_j-\delta^a_j\delta^b_i\right)
     \Delta\frac{3\sqrt{2}\pi}{g}
     \left(\frac{\mu^2}{2\pi^2}\right) .
\label{qq_cond}
\eea
We note that the superfluid density is sensitive to energies $p_0>\Delta$ 
and the energy dependence of the gap has to be kept. The color-favor
factor is 
\bea 
\left(\delta^a_i\delta^b_j-\delta^a_j\delta^b_i\right)
 \left(\frac{\vec\lambda}{2}\right)^{ac}
 \left(\frac{\vec\lambda}{2}\right)^{bd}  (M)_{ik}(M)_{jl}
 \left(\delta^c_k\delta^d_l-\delta^c_l\delta^d_k\right) 
 \hspace{1.25cm} && \nonumber \\
=-\frac{4}{3}\left\{  \Big( {\rm Tr}[M]\Big)^2 -{\rm Tr}\Big[ M^2\Big]
   \right\}\, . &&
\eea
We also note that the four-fermion operator is proportional to $g^2$ and 
the explicit dependence of the diagram on $g$ cancels. We find 
\cite{Son:1999cm,Schafer:2001za}
\be
\label{E_MM}
\Delta {\cal E} = -\frac{3\Delta^2}{4\pi^2}
 \left\{  \Big( {\rm Tr}[M]\Big)^2 -{\rm Tr}\Big[ M^2\Big]
   \right\}
 + \Big(M\leftrightarrow M^\dagger \Big).
\ee
This term can be matched against the $A_i$ terms in the effective
lagrangian. The result is \cite{Son:1999cm,Schafer:2001za}
\be
 A_1= -A_2 = \frac{3\Delta^2}{4\pi^2}, 
\hspace{1cm} A_3 = 0.
\ee

 We can now summarize the structure of the chiral expansion in the
CFL phase. The effective lagrangian has the form 
\be
{\mathcal L}\sim f_\pi^2\Delta^2 \left(\frac{\partial_0}{\Delta}\right)^k
 \left(\frac{\vec{\partial}}{\Delta}\right)^l
 \left(\frac{MM^\dagger}{p_F\Delta}\right)^m
 \left(\frac{MM}{p_F^2}\right)^n
  \big(\Sigma\big)^o\big(\Sigma^\dagger\big)^p.
\ee
Loop graphs in the effective theory are suppressed by powers of 
$\partial/(4\pi f_\pi)$. Since the pion decay constant scales as 
$f_\pi\sim p_F$ Goldstone boson loops are suppressed compared to 
higher order contact terms. We also note that the quark mass expansion 
is controlled by $m^2/(p_F\Delta)$. This is means that the chiral
expansion breaks down if $m^2\sim p_F\Delta$. This is the same
scale at which BCS calculations find a transition from the CFL 
phase to a less symmetric state. 

%%%%%%%%%%%%%%%%%%%%%%%%%%%%%%%%%%%%%%%%%%%%%%%%%%%%%%%%%%%%%%%%%%%%
\begin{figure}[t]
\bc\includegraphics[width=7cm]{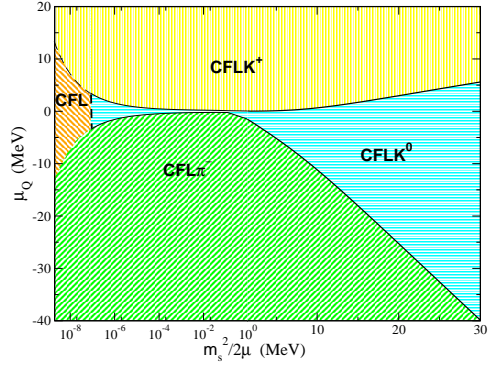}\ec
\caption{\label{fig_kcond}
Phase structure of CFL matter as a function of the effective 
chemical potential $\mu_s=m_s^2/(2p_F)$ and the lepton chemical 
potential $\mu_Q$, from Kaplan \& Reddy (2001). A typical value
of $\mu_s$ in a neutron star is 10 MeV. }
\end{figure}
%%%%%%%%%%%%%%%%%%%%%%%%%%%%%%%%%%%%%%%%%%%%%%%%%%%%%%%%%%%%%%%%%%%%

%%%%%%%%%%%%%%%%%%%%%%%%%%%%%%%%%%%%%%%%%%%%%%%%%%%%%%%%%%%%%%%%%%%%
\subsection{Kaon condensation}
\label{sec_kcond}
%%%%%%%%%%%%%%%%%%%%%%%%%%%%%%%%%%%%%%%%%%%%%%%%%%%%%%%%%%%%%%%%%%%%

 Using the chiral effective lagrangian we can determine the dependence 
of the order parameter on the quark masses. We will focus on the physically 
relevant case $m_s>m_u=m_d$. Because the main expansion parameter is 
$m_s^2/(p_F\Delta)$ increasing the quark mass is roughly equivalent to 
lowering the density. The effective potential for the order parameter is 
\be
\label{v_eff}
V_{eff} = \frac{f_\pi^2}{4} {\rm Tr}\left[
 2X_L\Sigma X_R\Sigma^\dagger -X_L^2-X_R^2\right] 
     - A_1\left[ \left({\rm Tr}(M\Sigma^\dagger)\right)^2 
     - {\rm Tr}\left((M\Sigma^\dagger)^2\right) \right].
\ee
The first term contains the effective chemical potential 
$\mu_s=m_s^2/(2p_F)$ and favors states with a deficit of 
strange quarks (with strangeness $S=-1$). The second term favors 
the neutral ground state $\Sigma=1$. The lightest excitation
with positive strangeness is the $K^0$ meson. We therefore
consider the ansatz $\Sigma = \exp(i\alpha\lambda_4)$ which
allows the order parameter to rotate in the $K^0$ direction. 
The vacuum energy is 
\be 
\label{k0+_V}
 V(\alpha) = -f_\pi^2 \left( \frac{1}{2}\left(\frac{m_s^2-m^2}{2p_F}
   \right)^2\sin(\alpha)^2 + (m_{K}^0)^2(\cos(\alpha)-1)
   \right),
\ee
where $(m_K^0)^2= (4A_1/f_\pi^2)m(m+m_s)$. Minimizing the vacuum 
energy we obtain 
\be 
\cos(\alpha)= \left\{ \begin{array}{cl}
 1 & \mu_s<m_K^0 \\
\;\frac{(m_K^0)^2}{\mu_s^2}\; & \mu_s >m_K^0\\
\end{array}\right.
\ee
The hypercharge density is 
\be 
n_Y = f_\pi^2 \mu_s \left( 1- \frac{(m_K^0)^4}{\mu_s^4}\right).
\ee
This result has the same structure as the charge density of a 
weakly interacting Bose condensate. Using the perturbative result 
for $A_1$ we can get an estimate of the critical strange quark mass.
We find  
\be
\label{ms_crit}
 m_s (crit)= 3.03\cdot  m_d^{1/3}\Delta^{2/3},
\ee
from which we obtain $m_s(crit)\simeq 70$ MeV for $\Delta\simeq 50$ 
MeV. This result suggests that strange quark matter at densities that 
can be achieved in neutron stars is kaon condensed. We also note that 
the difference in condensation energy between the CFL phase 
and the kaon condensed state is not necessarily small. For 
$\mu_s\to \Delta$ we have $\sin(\alpha)\to 1$ and $V(\alpha)\to 
f_\pi^2\Delta^2/2$. Since $f_\pi^2$ is of order $\mu^2/(2\pi^2)$
this is comparable to the condensation energy in the CFL phase. 

  The strange quark mass breaks the $SU(3)$ flavor symmetry to 
$SU(2)_I\times U(1)_Y$. In the kaon condensed phase this symmetry 
is spontaneously broken to $U(1)_Q$. If isospin is an exact symmetry 
there are two exactly massless Goldstone modes \cite{Schafer:2001bq},
the $K^0$ and the $K^+$. Isospin breaking leads to a small mass for 
the $K^+$. The phase structure as a function of the strange quark mass 
and non-zero lepton chemical potentials was studied by Kaplan and Reddy 
\cite{Kaplan:2001qk}, see Fig.~\ref{fig_kcond}. We observe that if the 
lepton chemical potential is non-zero charged kaon and pion condensates 
are also possible. 

%%%%%%%%%%%%%%%%%%%%%%%%%%%%%%%%%%%%%%%%%%%%%%%%%%%%%%%%%%%%%%%%%%%%
\begin{figure}[t]
\bc\includegraphics[width=9.5cm]{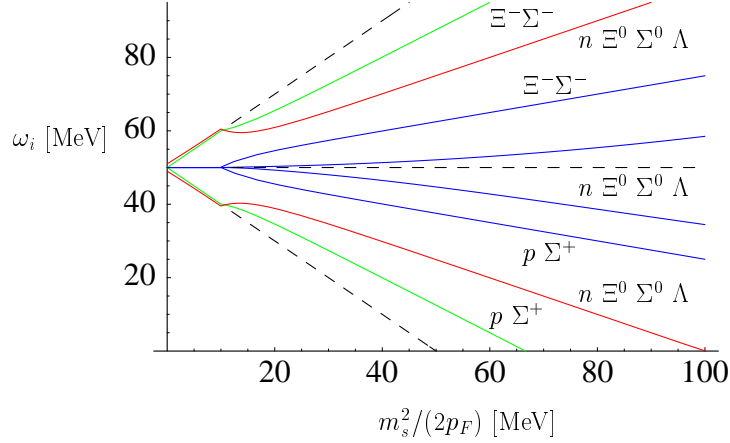}\ec
\caption{\label{fig_cfl_spec}
This figure shows the fermion spectrum in the CFL phase. For 
$m_s=0$ there are eight fermions with gap $\Delta$ and one
fermion with gap $2\Delta$ (not shown). Without kaon condensation
gapless fermion modes appear at $\mu_s=\Delta$ (dashed lines).
With kaon condensation gapless modes appear at $\mu_s=4\Delta/3$.}
\end{figure}
%%%%%%%%%%%%%%%%%%%%%%%%%%%%%%%%%%%%%%%%%%%%%%%%%%%%%%%%%%%%%%%%%%%%

%%%%%%%%%%%%%%%%%%%%%%%%%%%%%%%%%%%%%%%%%%%%%%%%%%%%%%%%%%%%%%%%%%
\subsection{Fermions in the CFL phase}
\label{sec_gCFL}
%%%%%%%%%%%%%%%%%%%%%%%%%%%%%%%%%%%%%%%%%%%%%%%%%%%%%%%%%%%%%%%%%%

 So far we have only studied Goldstone modes in the CFL phase.
However, as the strange quark mass is increased it is possible
that some of the fermion modes become light or even gapless
\cite{Alford:2003fq}. In order to study this question we 
have to include fermions in the effective field theory. 
The effective lagrangian for fermions in the CFL phase
is \cite{Kryjevski:2004jw,Kryjevski:2004kt}
\bea 
\label{l_bar}
{\mathcal L} &=&  
 {\rm Tr}\left(N^\dagger iv^\mu D_\mu N\right) 
 - D{\rm Tr} \left(N^\dagger v^\mu\gamma_5 
               \left\{ {\mathcal A}_\mu,N\right\}\right)
 - F{\rm Tr} \left(N^\dagger v^\mu\gamma_5 
               \left[ {\mathcal A}_\mu,N\right]\right)
  \nonumber \\
 & &  \mbox{} + \frac{\Delta}{2} \left\{ 
     \left( {\rm Tr}\left(N_LN_L \right) 
   - \left[ {\rm Tr}\left(N_L\right)\right]^2 \right)  
   - (L\leftrightarrow R) + h.c.  \right\}.
\eea
$N_{L,R}$ are left and right handed baryon fields in the 
adjoint representation of flavor $SU(3)$. The baryon fields 
originate from quark-hadron complementarity \cite{Schafer:1998ef}. 
We can think of $N$ as describing a quark which is surrounded 
by a diquark cloud, $N_L \sim q_L\langle q_L q_L\rangle$. The 
covariant derivative of the nucleon field is given by $D_\mu N
=\partial_\mu N +i[{\mathcal V}_\mu,N]$. The vector and axial-vector 
currents are 
\be
\label{v-av}
 {\mathcal V}_\mu = -\frac{i}{2}\left\{ 
  \xi \partial_\mu\xi^\dagger +  \xi^\dagger \partial_\mu \xi 
  \right\}, \hspace{1cm}
{\mathcal A}_\mu = -\frac{i}{2} \xi\left(\nabla_\mu 
    \Sigma^\dagger\right) \xi , 
\ee
where $\xi$ is defined by $\xi^2=\Sigma$. It follows that $\xi$ 
transforms as $\xi\to L\xi U(x)^\dagger=U(x)\xi R^\dagger$ with 
$U(x)\in SU(3)_V$. For pure $SU(3)$ flavor transformations $L=R=V$ 
we have $U(x)=V$. $F$ and $D$ are low energy constants that 
determine the baryon axial coupling. In perturbative QCD we
find $D=F=1/2$.

 The effective theory given in equ.~(\ref{l_bar}) can be derived 
from QCD in the weak coupling limit. However, the structure of the 
theory is completely determined by chiral symmetry, even if the 
coupling is not weak. In particular, there are no free parameters
in the baryon coupling to the vector current. Mass terms are also
strongly constrained by chiral symmetry. The effective chemical 
potentials $(X_L,X_R)$ appear as left and right-handed gauge 
potentials in the covariant derivative of the nucleon field.
We have 
\bea
\label{V_X}
 D_0N     &=& \partial_0 N+i[\Gamma_0,N], \\
 \Gamma_0 &=& -\frac{i}{2}\left\{ 
  \xi \left(\partial_0+ iX_R\right)\xi^\dagger + 
  \xi^\dagger \left(\partial_0+iX_L\right) \xi 
  \right\}, \nonumber 
\eea
where $X_L=MM^\dagger/(2p_F)$ and $X_R=M^\dagger M/(2p_F)$ as before.
$(X_L,X_R)$ covariant derivatives also appears in the axial vector 
current given in equ.~(\ref{v-av}).

 We can now study how the fermion spectrum depends on the quark mass.
In the CFL state we have $\xi=1$. For $\mu_s=0$ the baryon octet
has an energy gap $\Delta$ and the singlet has gap $2\Delta$. The 
correction to this result comes from the term
\be
{\rm Tr}\left(N^\dagger [\hat{\mu}_s,N])\right) 
 =\frac{\mu_s}{2} \left( 
  (\Xi^-)^\dagger(\Xi^-)+ (\Xi^0)^\dagger(\Xi^0)
   -(p)^\dagger(p)- (n)^\dagger(n)\right),
\ee
where $\hat{\mu}_s=\mu_s {\rm diag}(0,0,1)$. We observe that the excitation 
energy of the proton and neutron is lowered, $\omega_{p,n}=\Delta-\mu_s$, 
while the energy of the cascade states $\Xi^-,\Xi^0$ particles is raised, 
$\omega_{\Xi}=\Delta+\mu_s$. All other excitation energies are independent 
of $\mu_s$. As a consequence we find gapless $(p,n)$ excitations at 
$\mu_s=\Delta$. 

  This result is also easily derived in microscopic models 
\cite{Alford:2004hz}. The EFT perspective is nevertheless useful. In 
microscopic models the shift of the non-strange modes arises from 
a color chemical potential which is needed in order to neutralize 
the system. The effective theory is formulated directly in terms 
of gauge invariant variables and no color chemical potentials are 
needed. The shift in the non-strange modes is due to the fact that 
the gauge invariant fermion fields transform according to the 
adjoint representation of flavor $SU(3)$.

  The situation is more complicated when kaon condensation is taken 
into account. In the kaon condensed phase there is mixing in the 
$(p,\Sigma^+,\Sigma^-,\Xi^-)$ and $(n,\Sigma^0,\Xi^0,\Lambda^8,
\Lambda^0)$ sector. For $m_K^0\ll\mu_s\ll \Delta$ the spectrum is 
given by
\be
\omega_{p\Sigma^\pm\Xi^-}= \left\{
 \begin{array}{c}
 \Delta \pm \frac{3}{4}\mu_s, \\
 \Delta \pm \frac{1}{4}\mu_s,
\end{array}\right.  \hspace{0.75cm}
\omega_{n\Sigma^0\Xi^0\Lambda} = \left\{
 \begin{array}{c}
   \Delta \pm \frac{1}{2}\mu_s ,\\ 
   \Delta , \\
   2\Delta .
 \end{array} \right. 
\ee
Numerical results for the eigenvalues are shown in Fig.~\ref{fig_cfl_spec}. 
We observe that mixing within the charged and neutral baryon sectors leads 
to level repulsion. There are two modes that become light in the CFL window 
$\mu_s\leq 2\Delta$. One mode is a linear combination of the proton and 
$\Sigma^+$ and the other mode is a linear combination of the neutral 
baryons $(n,\Sigma^0,\Xi^0,\Lambda^8,\Lambda^0)$.

%%%%%%%%%%%%%%%%%%%%%%%%%%%%%%%%%%%%%%%%%%%%%%%%%%%%%%%%%%%
\begin{figure}[t]
\bc\includegraphics[width=6.3cm]{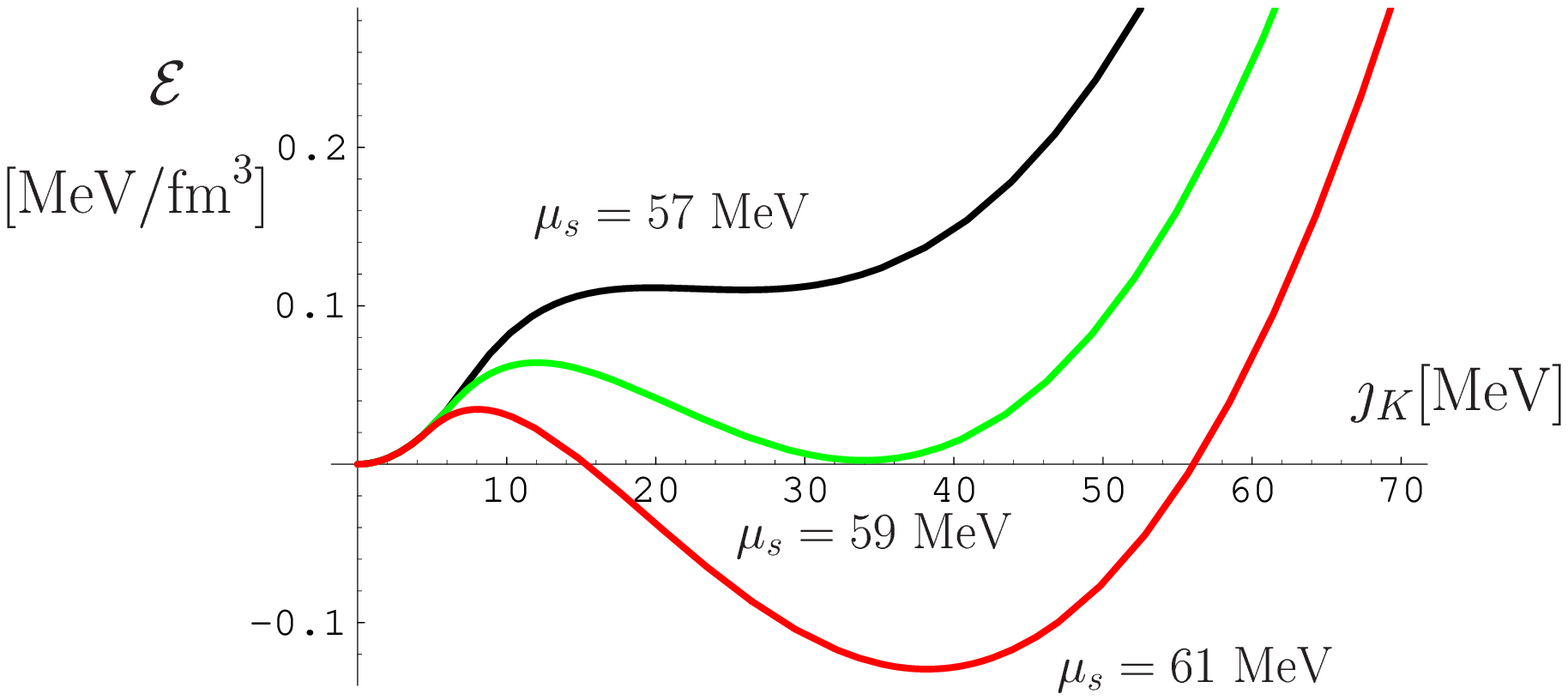}\hspace*{-0.3cm}
\includegraphics[width=6.3cm]{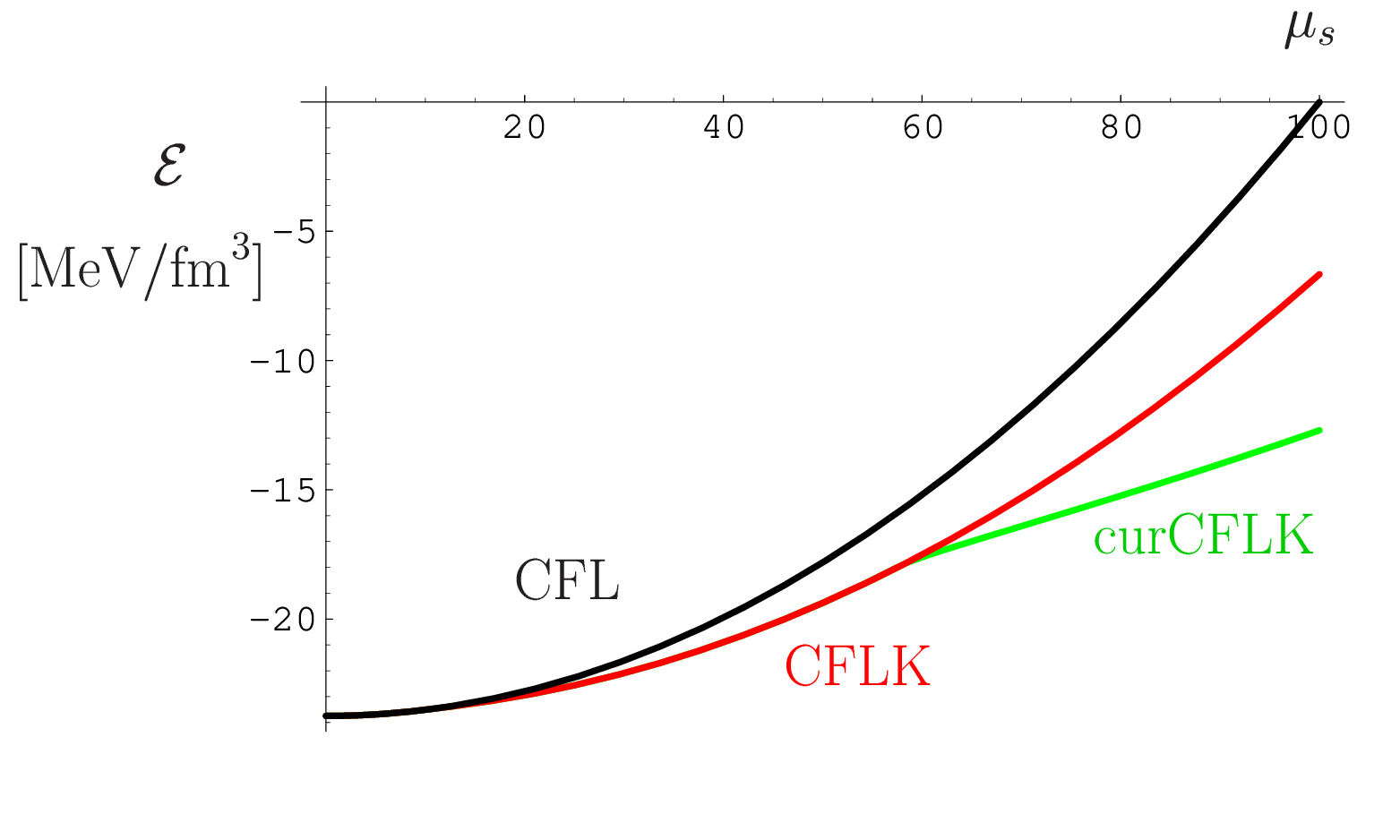}\ec
\caption{Left panel: Energy density as a function of the current 
$\jmath_K$ for several different values of $\mu_s=m_s^2/(2p_F)$ 
close to the phase transition. Right panel: Ground state energy 
density as a function of $\mu_s$. We show the CFL phase, the 
kaon condensed CFL (KCFL) phase, and the supercurrent state
(curKCFL).}
\label{fig_jfct}
\end{figure}
%%%%%%%%%%%%%%%%%%%%%%%%%%%%%%%%%%%%%%%%%%%%%%%%%%%%%%%%%%%

%%%%%%%%%%%%%%%%%%%%%%%%%%%%%%%%%%%%%%%%%%%%%%%%%%%%%%%%%%%%%%%%%
\subsection{Meson supercurrent state}
\label{sec_cur}
%%%%%%%%%%%%%%%%%%%%%%%%%%%%%%%%%%%%%%%%%%%%%%%%%%%%%%%%%%%%%%%%%

 Recently, several groups have shown that gapless fermion modes lead 
to instabilities in the current-current correlation function 
\cite{Huang:2004bg,Casalbuoni:2004tb}. Motivated by these results we 
have examined the stability of the kaon condensed phase against the 
formation of a non-zero current \cite{Schafer:2005ym,Kryjevski:2005qq}.
Consider a spatially varying $U(1)_Y$ rotation of the maximal kaon 
condensate
\be 
U(x)\xi_{K^0} U^\dagger (x) = \left(
 \begin{array}{ccc}
 1 & 0 & 0 \\
 0 & 1/\sqrt{2} & ie^{i\phi_K(x)}/\sqrt{2} \\
 0 & ie^{-i\phi_K(x)}/\sqrt{2} & 1/\sqrt{2} 
\end{array} \right).
\ee
This state is characterized by non-zero currents
\be
\label{cur}
\vec{\cal V} =  \frac{1}{2}\left(\vec{\nabla} \phi_K\right) \left(
 \begin{array}{ccc}
0 & 0 & 0 \\
0 & 1 & 0 \\
0 & 0 & -1 
\end{array} \right),\hspace{0.3cm}
\vec{\cal A} =  \frac{1}{2}\left(\vec{\nabla} \phi_K\right) \left(
 \begin{array}{ccc}
0 & 0 & 0 \\
0 & 0 & -ie^{i\phi_K} \\
0 & ie^{-i\phi_K} & 0 
\end{array} \right).
\ee
In the following we compute the vacuum energy as a function 
of the kaon current $\vec{\jmath}_K=\vec\nabla\phi_K$. The meson 
part of the effective lagrangian gives a positive contribution
\be
{\cal E}=\frac{1}{2}v_\pi^2f_\pi^2\jmath_K^2 .
\ee
A negative contribution can arise from gapless fermions. In order 
to determine this contribution we have to calculate the fermion spectrum 
in the presence of a non-zero current. The relevant part of the
effective lagrangian is  
\bea 
{\cal L} &=& {\rm Tr}\left(N^\dagger iv^\mu D_\mu N\right)
 + {\rm Tr}\left(N^\dagger \gamma_5 \left( \rho_A+\vec{v}\cdot
      \vec{\cal A}\right) N\right) \nonumber \\
 & & \mbox{}  
 +\frac{\Delta}{2} \left\{ {\rm Tr}\left(N N\right) -
  {\rm Tr}\left(N\right){\rm Tr}\left(N\right)+ h.c.\right\},
\eea
where we have used $D=F=1/2$. The covariant derivative is 
$D_0N=\partial_0N+i[\rho_V,N]$ and $D_iN=\partial_i N +i
\vec{v}\cdot[\vec{\cal V},N]$ with $\vec{\cal V},\vec{\cal A}$ given in 
equ.~(\ref{cur}) and  
\be 
\rho_{V,A} = \frac{1}{2}\left\{ 
  \xi \frac{M^\dagger M}{2p_F}\xi^\dagger \pm 
  \xi^\dagger \frac{MM^\dagger}{2p_F} \xi 
  \right\}.
\ee
The vector potential $\rho_V$ and the vector current $\vec{\cal V}$
are diagonal in flavor space while the axial potential $\rho_A$ 
and the axial current $\vec{\cal A}$ lead to mixing. The fermion 
spectrum is quite complicated. The dispersion relation of the lowest 
mode is approximately given by
\be
\label{disp_ax}
\omega_l = \Delta +\frac{(l-l_0)^2}{2\Delta}-\frac{3}{4}
  \mu_s -\frac{1}{4}\vec{v}\cdot\vec{\jmath}_K,
\ee
where $l=\vec{v}\cdot\vec{p}-p_F$ and we have expanded $\omega_l$ 
near its minimum $l_0=(\mu_s+\vec{v}\cdot\vec{\jmath}_K)/4$.
Equation (\ref{disp_ax}) shows that there is a gapless mode if 
$\mu_s>4\Delta/3-\jmath_K/3 $. The contribution of the gapless mode 
to the vacuum energy is 
\be
\label{e_fct} 
{\cal E} = \frac{\mu^2}{\pi^2}\int dl \int 
 \frac{d\Omega}{4\pi} \;\omega_l \theta(-\omega_l) ,
\ee
where $d\Omega$ is an integral over the Fermi surface. The integral 
in equ.~(\ref{e_fct}) receives contributions from one of the pole caps 
on the Fermi surface. The result has exactly the same structure as the 
energy functional of a non-relativistic two-component Fermi liquid with 
non-zero polarization, see \cite{Son:2005qx}. Introducing dimensionless 
variables 
\be 
\label{x+h}
 x = \frac{\jmath_K}{a\Delta}, \hspace{0.5cm}
 h = \frac{3\mu_s-4\Delta}{a\Delta}.
\ee
we can write ${\cal E} = c\, {\cal N} f_h(x)$ with
\be
\label{cur_fct}
 f_h(x) = x^2-\frac{1}{x}\left[
   (h+x)^{5/2}\Theta(h+x) - (h-x)^{5/2}\Theta(h-x) \right] .
\ee
We have defined the constants
\be
\label{consts}
 c = \frac{2}{15^4c_\pi^3 v_\pi^6},\hspace{0.45cm}
 {\cal N} =   \frac{\mu^2\Delta^2}{\pi^2},\hspace{0.45cm}
 a = \frac{2}{15^2 c_\pi^2 v_\pi^4}, 
\ee
where $c_\pi = (21-8\log(2))/36$ is the numerical coefficient 
that appears in the weak coupling result for $f_\pi$. 
According to the analysis in \cite{Son:2005qx} the function
$f_h(x)$ develops a non-trivial minimum if $h_1<h<h_2$ with 
$h_1\simeq -0.067$ and $h_2\simeq 0.502$. In perturbation 
theory we find $a=0.43$ and the kaon condensed ground state
becomes unstable for $(\Delta- 3\mu_s/4) < 0.007\Delta$. 

 The energy density as a function of the current and the 
groundstate energy density as a function of $\mu_s$ are 
shown in Fig.~\ref{fig_jfct}. In these plots we have included 
the contribution of a baryon current $\jmath_B$, as suggested 
in \cite{Kryjevski:2005qq}. In this case we have to minimize the 
energy with respect to two currents. The solution is of the form
$\jmath_B\sim \jmath_K$. The figure shows the dependence on 
$\jmath_K$ for the optimum value of $\jmath_B$. We have not properly 
implemented electric charge neutrality. Since the gapless mode is 
charged, enforcing electric neutrality will significantly suppress 
the magnitude of the current. We have also 
not included the possibility that the neutral mode becomes gapless. 
This will happen at somewhat larger values of $\mu_s$.

 We note that the ground state has no net current. This is clear 
from the fact that the ground state satisfies $\delta {\cal E}/\delta 
(\vec\nabla\phi_K)=0$. As a consequence the meson current is canceled 
by an equal but opposite contribution from gapless fermions. We also 
expect that the ground state has no chromomagnetic instabilities. 
The kaon current is equivalent to an external gauge field. By 
minimizing the thermodynamic potential with respect to $\jmath$
we ensure that the second derivative, and therefore the screening 
mass, is positive.

%%%%%%%%%%%%%%%%%%%%%%%%%%%%%%%%%%%%%%%%%%%%%%%%%%%%%%%%%%%%%%%%%%
\section{Conclusion: The many uses of effective field theory}
\label{sec_sum}
%%%%%%%%%%%%%%%%%%%%%%%%%%%%%%%%%%%%%%%%%%%%%%%%%%%%%%%%%%%%%%%%%%

 Strongly correlated quantum many body systems play a role in 
many different branches of physics, atomic physics, condensed
matter physics, nuclear and particle physics. One of the main 
themes of these lectures is the idea that effective field theories
provide a unified description of systems that involve vastly 
different scales. For example, nuclear physicists studying 
neutron matter have learned a great deal from studying cold
atomic gases (and vice versa). Similarly, progress in understanding 
non-Fermi liquid behavior in strongly correlated electronic systems
has been helpful in understanding dense quark matter in QCD. It
is our hope that these lecture notes will play a small part in 
fostering exchange of ideas between different communities in 
the future. 

 Acknowledgments: I would like to thank the organizers of the 
Trento school, Janos Polonyi and Achim Schwenk, for doing such 
an excellent job in putting the school together, and all the 
students who attended the school for turning it into a stimulating 
experience. This work was supported in part by US DOE grant 
DE-FG02-03ER41260.

%\newpage

\printindex
\end{document}